\documentclass[aps,eqsecnum,nofootinbib,superscriptaddress,showpacs]{revtex4}
\usepackage{amsmath}
\usepackage{amsfonts}
\usepackage{amssymb}
\usepackage[dvips]{graphicx}  % for graphics
\usepackage{amsmath,amsfonts}
\usepackage{color}  % for color
\def\be{\begin{eqnarray}}
\def\ee{\end{eqnarray}}
\def\beq{\begin{equation}}
\def\eeq{\end{equation}}

\def\({\left (}
\def\){\right )}
\def\S{{\cal S}}

\def\tr{\mathrm{tr}}

\newtheorem{theorem}{Theorem}[section]
\newtheorem{lemma}[theorem]{Lemma}
\newtheorem{proposition}[theorem]{Proposition}
\newtheorem{corollary}[theorem]{Corollary}
\newtheorem{definition}{Definition}[section]
\newtheorem{remark}[theorem]{Remark}

\newcommand{\qed}{\nobreak \ifvmode \relax \else
      \ifdim\lastskip<1.5em \hskip-\lastskip
      \hskip1.5em plus0em minus0.5em \fi \nobreak
      \vrule height0.75em width0.5em depth0.25em\fi}
\newcommand{\qcd}{\begin{flushright} $\Box$ \end{flushright}}
\begin{document}
%%%%%%%%%%%%%%%%%%%%%%%%%%%%%%%%%%%%%%%%%%%%%%%%%%%%%%%%%%%%%
%%%%%%%%%%%%%%%%%%%%%%   Title Page   %%%%%%%%%%%%%%%%%%%%%%%
%\rightline{Preprint-MTM}
%\rightline{hep-th/0605224}

\title{
Lightlike sets with applications to the rigidity of null geodesic incompleteness
}

\author{I.P. Costa e Silva}
\email{pontual.ivan@ufsc.br}
\affiliation{Department of Mathematics,\\
Universidade Federal de Santa Catarina \\88.040-900 Florian\'{o}polis-SC, Brazil}
\author{J.L. Flores}
\email{floresj@uma.es}
\affiliation{Departamento de \'Algebra, Geometr\'{i}a y Topolog\'{i}a,\\ Facultad de Ciencias, Universidad de M\'alaga \\ Campus Teatinos, 29071 M\'alaga, Spain}

\date{\today}

\begin{abstract}
An important, if relatively less well known aspect of the singularity theorems in Lorentzian Geometry is to understand how their conclusions fare upon weakening or suppression of one or more of their hypotheses. Then, theorems with modified conclusion may arise, showing that those conclusions will fail only in special cases, at least some of which may be described. These are the so-called {\em rigidity theorems}, and have many important examples in the specialized literature. In this paper, we prove rigidity results for generalized plane waves and certain globally hyperbolic spacetimes in the presence of maximal compact surfaces. Motivated by some general properties appearing in these proofs, we develop the theory of {\em lightlike sets}, entities similar to achronal sets, but more appropriate to deal with low-regularity null submanifolds.

\end{abstract}
%
%\pacs{04.20.Dw;04.20.Gz;02.40.-k}
%
\maketitle

\section{Introduction} \label{0}

One of the most distinctive aspects of Lorentzian Geometry is its emphasis on obtaining well motivated sufficient conditions ensuring that spacetimes have incomplete nonspacelike geodesics, via the many singularity theorems available in the literature. This is only to be expected, given the key importance of singularities in describing gravitational collapse at the core of black holes or the fate of the Universe in geometric theories of gravity.

However, an equally important aspect of singularity theorems (and indeed of any theorem with physical applicability) is to understand the naturalness of their hypotheses, both from mathematical and physical standpoints. On the one hand, from a physical perspective, one would wish to retain only those assumptions which are likely to have an adequate counterpart in Nature. On the other hand, this should be reflected mathematically in that exceptions to the conclusions of a theorem, upon weakening or altogether suppressing one or more of its assumptions, should only happen in special (in a suitable sense) circumstances. When this does occur, one refers to the result as {\em rigid}, and to the modified assertions arising thereby as {\em rigidity (or rigid) theorems}. Theorems of this sort have long been known and pursued in the mathematical literature. A beautiful description of the general philosophy behind rigidity theorems, together with a number of key specific examples germane to the context of Mathematical Relativity, can be found in the Chapter 14 of Ref. \cite{BE}.

Now, in the classic Penrose singularity theorem \cite{oneill,Penrose}, one assumes the existence of a closed trapped surface $\Sigma$ in spacetime. The condition that $\Sigma$ be a closed trapped surface was Penrose's geometric surrogate of ``critical matter density'' having been reached in its neighbourhood, signalling an impending gravitational collapse. Mathematically, this condition means (under suitable orientability assumptions) that the so-called {\em null mean curvatures (or expansion scalars)} $\theta_{\pm}: \Sigma \rightarrow \mathbb{R}$, which measure the initial divergence of the two families of normal null geodesics emanating from $\Sigma$, are both strictly negative, i.e., $\theta_{\pm} <0$. With additional physically natural assumptions, the null geodesic incompleteness of spacetime then follows.

Natural as this scheme may seem, however, rigorous proofs of the existence of trapped surfaces in the presence of large concentrations of matter can be very difficult to come by in spacetimes without special symmetries. Much more natural entities whose existence has been proven in a number of physically motivated contexts \cite{AEM,AMMS,AM,eichmair1,eichmair2,SY3} are marginally outer trapped surfaces (MOTS), for which one has $\theta_{+} \equiv 0$. MOTS are naturally related to closed trapped surfaces, since they often appear as boundaries of compact spatial regions in spacetime containing closed trapped surfaces and as spatial sections of stationary black hole horizons.

A natural question is then: how does geodesic incompleteness fare when one requires that the spacetime contains a MOTS instead of a closed trapped surface, say in Penrose's singularity theorem? Of course, simple examples obtained by making identifications in Minkowski spacetime show that one need not have geodesic incompleteness in this case. However, it has recently been shown \cite{EGP,me2,CG} that under reasonable additional assumptions, spacetimes containing MOTS are {\em generically} geodesically incomplete (in a precise sense - see especially \cite{CG}). In other words, geodesic incompleteness fails only in ``special situations'', being therefore rigid in the sense described above. This, in turn, leads one to look for rigidity theorems describing the exceptional cases. The typical setting of such theorems is to assume that spacetime is (timelike or null) geodesically complete and then deduce (under some other natural assumptions) special features of spacetime.

In this paper, we prove rigidity theorems involving {\em maximal} submanifolds, i.e., those having zero mean curvature. These are closely related to MOTS, and also appear in various natural situations. Indeed, the model for our basic rigidity result (cf. Theorem \ref{main2} below) is Theorem 7.1 of Ref. \cite{EGP}, which is a rigidity result for MOTS, and our proof therein is an adaptation of the proof in \cite{EGP}. Theorem \ref{main2} will in turn serve as a basis to obtain a few other rigidity results, which provide descriptions of certain classes of geodesically complete spacetimes of physical interest. In addition, most of our results remain valid for any spacetime dimension larger than two without assuming any field equations.

The general properties of null geodesics and achronal sets appearing in the proofs have led us to investigate certain subsets of spacetime of independent interest, which we have christened {\em lightlike sets}. These sets generalize a concept first introduced by G. Galloway in \cite{gallowaymaximum} (cf. Definition 3.1 of that reference).

The rest of the paper is organized as follows.

In Section \ref{0.1} we merely set the general conventions, nomenclature and notation for our main results. This section is accordingly very short.

In Section \ref{1} we define lightlike sets (Definition \ref{def1.1}) and investigate in detail their basic properties. Lightlike sets have properties analogous to those of achronal sets in the causal theory of spacetimes. So, we have found it worthwhile to develop their theory in a general fashion, roughly having the abstract theory of achronal sets and boundaries as models. Accordingly, not all of the results on lightlike sets will find a direct application in later sections. The reader more interested in the rigidity theorems may therefore just skim over this section in a first reading, referring back to it afterwards as needed.

Sections \ref{2} and \ref{3} mostly review some relatively well known results in order to keep the paper relatively self-contained. However, towards the end of Section \ref{3} we prove a basic rigidity result (Theorem \ref{main2}) of which we make extensive use in later sections.

Section \ref{4} presents a rigidity theorem (Theorem \ref{main3}) for {\em generalized plane waves}, together with a few corollaries. These comprise a very well studied class of spacetimes \cite{CFSgrg} which include important exact solutions of the Einstein field equation in General Relativity (\cite{Brinkmann,ER,Yurtsever}). The motivation to consider this particular class of spacetimes has arisen from a conjecture by J. Ehlers and K. Kundt \cite{EK}, stating that every geodesically complete vacumm (i.e., Ricci flat) pp-wave spacetime is a plane-wave spacetime (cf. Ch. 13 of \cite{BE}). If that conjecture is true, the in particular that the spacetime is, in addition to the other assumptions, globally hyperbolic, then it is Minkowski spacetime. Theorem \ref{main3} and its corollaries can be viewed as settling in the affirmative some particular cases of this conjecture.

Section \ref{5} establishes (cf. Theorem \ref{bifurcate}) the existence of a bifurcate horizon structure in certain globally hyperbolic spacetimes without assuming the existence any Killing vector fields. We also present therein a concrete context (cf. Theorem \ref{reasonable}) in which these bifurcate horizons do occur.

%%%%%%%%%%%%%%%%%%%%%%%%%%%%%%%%%%%%%%%%%%%%%%%%%%%%%%%%%%%%%
\section{Notation \& Conventions}\label{0.1}
%%%%%%%%%%%%%%%%%%%%%%%%%%%%%%%%%%%%%%%%%%%%%%%%%%%%%%%%%%%%%

 In all that follows, we fix a {\em spacetime}, i.e, an ($n+1$)-dimensional ($n \geq 2$), second-countable, connected, Hausdorff, smooth (i.e., $C^{\infty}$) Lorentz manifold $M$ endowed with a smooth metric tensor $g$ (signature $(-,+, \ldots,+)$), and with a fixed time-orientation. We assume that the reader is familiar with the basic definitions and results of global Lorentzian Geometry and the causal theory of spacetimes, as found in the standard references \cite{BE,HE,oneill}. All submanifolds of $M$ are regarded as $C^{\infty}$, embedded, and their topology is the induced topology. Finally, we follow the convention that nonspacelike vectors are always nonzero, and we use the terms ``nonspacelike'' and ``causal'' interchangeably. If there is no risk of confusion, we shall often use the notation $\langle \,. , \, . \rangle = g( \,. , \, . )$.

In order to set further conventions for the results of this paper, let $N \subseteq M$ be a semi-Riemannian submanifold. Recall that the covariant derivative on $(M,g)$ computed for vectors fields $X,Y \in \mathfrak{X}(N)$ decomposes in a unique fashion as
\[
\nabla ^M_X Y = \nabla ^{N}_X Y + \mathfrak{S}(X,Y),
\]
where the first term on the right hand side is the covariant derivative wrt the induced metric (which is either Lorentzian or Riemannian in the present context) on $N$, and the second term defines a normal bundle-valued symmetric tensor called the {\em shape tensor}, or {\em second fundamental form tensor} of $N \hookrightarrow M$. Given $p \in N$ and $\{e_1, \ldots, e_{\mbox{dim }N}\} \subset T_p N$ an orthonormal basis, the {\em mean curvature vector} $H_p$ at $p$ is the trace of the shape tensor
\[
H_p = \mbox{tr}_{N} \; \mathfrak{S} = \sum_{i=1}^{\mbox{dim }N}\mathfrak{S}(e_i,e_i).
\]
(This trace does not depend on the orthonormal basis chosen). $N$ is said to be {\em totally geodesic} if $\mathfrak{S} \equiv 0$ and {\em maximal} if $H_p \equiv 0$, $\forall p \in \Sigma$.
It can be shown (cf. Proposition 13, p. 104 of Ref. \cite{oneill}) that $N$ is totally geodesic iff any geodesic starting at a point of $N$ and tangent to it remains initially in $N$.

By a {\em surface} we will always mean a connected, acausal (hence spacelike), codimension $2$ submanifold $\Sigma \subset M$. We say that the surface $\Sigma$ is {\em two-sided} if its normal bundle $N\Sigma$ is trivial. In this case, we can pick two linearly independent, normal future-directed null vector fields $K_{\pm} : \Sigma \rightarrow N \Sigma$ and define the {\em null mean curvatures} $\theta_{\pm} \in C^{\infty}(\Sigma)$ of $\Sigma$ by
\[
\theta_{\pm}(p) = - \langle H_p, K_{\pm}(p) \rangle ,
\]
for all $p \in \Sigma$. By these conventions, an ordinary sphere in a spatial ``$t=0$'' section of Minkowski spacetime has $\theta_{+} \theta_{-} <0$. $\Sigma$ is a closed trapped surface (resp. a MOTS) if it is compact, two-sided and $\theta_{\pm} <0$ (resp. $\theta_{+} = 0$). Since the vector fields $K_{\pm}$ are unique up to rescaling by positive smooth functions, such (in)equalities do not depend on their particular choice. Of course, every compact two-sided maximal surface in $(M,g)$ is a MOTS, but the converse does not necessarily hold.

Given any set $S \subset M$, let $\eta:[0,a) \rightarrow M$ be a nonspacelike curve starting at $S$. We say that $\eta$ is an {\em $S$-ray} if the Lorentzian length of any segment of $\eta$ starting at $S$ up to any point $p$ along the curve realizes the Lorentzian distance $d(S, p) \equiv \mbox{sup} _{q \in  S} d(q,p)$ from $S$ to that point, where $d$ denotes the Lorentzian distance function (cf. Definition 14.4 of \cite{BE}). It is easy to check that such an $S$-ray will have a reparametrization as a geodesic. In this paper, we will mostly consider null $\Sigma$-rays emanating from a surface $\Sigma$, in which case they have globally achronal images, are normal to $\Sigma$ and have no focal points.

\section{Lightlike sets} \label{1}

As mentioned in the Introduction, in this section we develop the basics of a general theory of lighlike sets.

\begin{definition}
\label{def1.1}
A non-empty set $A \subseteq M$ is said to be {\em future} [resp. {\em past}] {\em lightlike} if
\begin{itemize}
\item[i)] $A$ is {\em locally achronal}, i.e., $\forall x \in A, \exists U \ni x$ open such that $A \cap U$ is achronal in $(U,g|_{U})$;
\item[ii)] For each $x \in A$ and for each $U \ni x$ neighbourhood of $x$ for which $A \cap U$ is achronal in $(U, g|_{U})$, there exists $y \in A \cap U$ such that $y \in J^{+}(x,U)\setminus \{x\}$ [resp. $y \in J^{-}(x,U)\setminus \{x\}$].
\end{itemize}
A non-empty set is {\em lightlike} if it is either past or future lightlike, or both.
\end{definition}

The following four propositions provide some important examples of lightlike sets.

\begin{proposition}
\label{Ex1.1}
Let $\alpha: [a,b) \rightarrow M$ be a future-directed null geodesic ($-\infty < a < b \leq + \infty $), and $A \subset \mbox{Im} \alpha$ be any set dense in the image of $\alpha$, i.e., such that $\mbox{Im} \alpha \subseteq \overline{A}$. Suppose that $(M,g)$ is strongly causal. Then, $A$ is a future lightlike set.
\end{proposition}
{\em Proof.} Pick any $t_0 \in [a,b)$ with $x_0 = \alpha(t_0) \in A$. Let $V \ni x_0$ be a convex normal neighbourhood and (using strong causality) pick any $U \ni x_0$ neighbourhood of $x_0$ such that $U \subseteq V$ and for any future-directed causal curve $\beta:[0,1] \rightarrow M$ with $\beta(0),\beta(1) \in U$, we have that $\beta[0,1] \subset V$.

Let $x,y \in U \cap A$, $x \neq y$. We can assume that there exist $t,s \in [a,b)$, $t < s$, with, say, $x = \alpha(t), y = \alpha(s)$. Then $\alpha[t,s] \subset V$ from our choice of $U$. Since $V$ is convex normal, $\alpha|_{[t,s]}$ is maximal in $(V, g|_{V})$, and hence its image is achronal therein. In particular, $ x \not \ll _{U} y$. Thus, $U \cap A$ is achronal in $(U, g|_{U})$. This proves that $A$ is locally achronal.

Now, take $x_0\in A$ and let $U\ni x_0$ be any neighbourhood for which $A \cap U$ is achronal in $(U, g|_{U})$. By continuity, we can pick a number $\epsilon >0$ such that $t_0+ \epsilon \in [a,b)$ and for which $\alpha[t_0, t_0 + \epsilon ] \subset U$. Since $(M,g)$ is causal, $\alpha(t_0) \neq \alpha(t_0 + \epsilon)$. $M$ is Hausdorff, so there exist $V_1,V_2 \subset U$ disjoint neighbourhoods of $\alpha(t_0)$ and $\alpha(t_0 + \epsilon)$, respectively. Since $\alpha(t_0 + \epsilon) \in \overline{A}$, for some $t_0 < t \leq t_0 + \epsilon$, $\alpha(t) \in A \cap V_2$, and in particular $\alpha(t) \in J^{+}(x_0,U) \setminus \{x_0 \}$.
\qcd

Recall that {\em achronal boundaries} are sets of the form $\partial I^{\pm}(C)$ for some set $C \subseteq M$ \cite{P}. In general, achronal boundaries are {\em not} lightlike sets. For instance, consider the Lorentzian cylinder $(M,g)$ obtained from the 2-dimensional Minkowski spacetime by an isometric identification along a spatial direction, and pick any $p \in M$ thereon. The achronal boundaries $\partial I^{\pm}(p)$ are neither future nor past lightlike, since condition (ii) in Definition \ref{def1.1} fails to hold. But in any case we have the following result.

\begin{proposition}
\label{Ex1.2}
Let $C \subseteq M$ be any set. Then, $A := \partial I^{-}(C)\setminus \overline{C}$, if non-empty, is an achronal future lightlike set.
\end{proposition}
{\em Proof.} From standard properties of achronal boundaries (see, e.g., \cite{oneill,P}), $A$ is an achronal (hence locally achronal) $C^0$ hypersurface, and given any point $x \in  \partial I^{-}(C)\setminus \overline{C}$, there exists a future-directed null geodesic $\alpha: [0,a) \rightarrow M$ such that $\alpha(0) =x$, $\alpha[0,a) \subset \partial I^{-}(C)\setminus \overline{C}$, and which is either future-inextendible, or has a future endpoint on $\overline{C}$. Thus, given any open set $U \ni x$ such that $U \cap \overline{C} = \emptyset$, by continuity there exists a number $0 < t_0 < a$ such that $\alpha[0,t_0] \subset U \cap A$ and $\alpha(t_0) \in J^{+}(x,U) \setminus \{x\}$.
\qcd

\begin{proposition}
\label{Ex1.3}
Let $S \subseteq M$ be any non-empty set and  $\alpha: [0,a) \rightarrow M$ be a future-directed null $S$-ray. Then $A := \alpha[0,a)$ is an achronal future lightlike set.
\end{proposition}
{\em Proof.} Since every segment of $\alpha$ is maximal, $A$ is achronal. For each $t_0 \in [0,a)$, and each neighbourhood $U$ of $x_0=\alpha(t_0)$, we can pick, by continuity, $t_0 < t_1 < a$ with $\alpha[t_0,t_1] \subset U$. We can also assume that $\alpha(t_0) \neq \alpha(t_1)$, since $\alpha|_{[t_0,t_1]}$ cannot be constant (recall that null vectors are nonzero according to our conventions). So $\alpha(t_1) \in J^{+}(x_0,U) \setminus \{x_0\}$.
\qcd

It is easy to see that each of these examples has a time-dual version which is a {\em past} lightlike set. The following proposition provides an example of a lightlike set.

\begin{proposition}
\label{prop1.2}
If $S \subset M$ is a locally achronal null submanifold, then it is a (future and past) lightlike set.
\end{proposition}
{\em Proof.} It is well known (see, e.g., Prop. 4 of \cite{kupeli}) that $S$ admits a future-directed, everywhere non-zero null tangent vector field $K: S \rightarrow TS$, unique up to rescaling. Since $S$ is locally achronal, all that remains to be shown is clause (ii) in Definition \ref{def1.1}. Pick any $p \in S$ and a neighbourhood $U \ni p$ in $M$ for which $U \cap S$ is achronal in $(U, g|_U)$. We can then consider an integral curve $\alpha: (-\epsilon, \epsilon) \rightarrow S \cap U$ of $K$ with $\alpha '(0) = K_p$. By choosing $\epsilon >0 $ small enough, we can assume that $\alpha$ is one-to-one, and clause (ii) then follows.
\qcd

As we have seen, a lightlike set need not be a manifold. However, the following proposition, which can be seen as a partial converse to Proposition \ref{prop1.2}, states that if it is a manifold, then it must be a {\em null} (a.k.a. lightlike) submanifold (i.e. an embedded $C^{\infty}$ submanifold with everywhere degenerate induced metric), thus motivating its name.

\begin{proposition}
\label{prop1.1}
If a lightlike set $S \subset M$ is a submanifold, then it is a null submanifold.
\end{proposition}
{\em Proof.} Assume, for definiteness, that $S$ is future lightlike. (The proof when it is past lightlike is analogous.) Suppose that for some $p \in S$, the tangent space $T_pS$ is not a degenerate subspace of $T_pM$. Then it is either timelike, or spacelike. Suppose that $T_pS$ is timelike, and hence it contains a timelike vector $v$, say. In this case, there exists a timelike curve $\alpha: (-\epsilon, \epsilon) \rightarrow S$ with $\alpha '(0) = v$, in contradiction with the local achronality of $S$. Thus, we can suppose that $T_pS \subseteq T_pM$ is a spacelike subspace. Let $U$ be a convex normal neighborhood of $p$ in $M$ for which $U \cap S$ is an achronal spacelike submanifold in $(U, g|_U)$, and let $q\in S\cap U$ such that $q\in J^+(p,U)\setminus \{p\}$. Then, there exists a null geodesic $\gamma$ in $U$ connecting $p$ with $q$. By redefining $U$ if necessary, we can also assume that $\tilde{S}=\exp_p^{-1}(S\cap U)$ is a spacelike submanifold of $T_p M$. Clearly, $S\cap U$ intersects $\gamma$ at least at $p,q\in U$. Therefore, $\tilde{S}$ also intersects $\tilde{\gamma}=\exp_p^{-1}(\gamma)$ at least at $\exp_p^{-1}(p)=0$ and $\tilde{q}=\exp_p^{-1}(q)$. But, we would have a spacelike submanifold $\tilde{S}$ of $T_p M$ touching $\tilde{\gamma}$ (which is part of the null cone of $T_p M$ at zero) in two diferent points $0$, $\tilde{q}$, which is absurd (note that $T_p M$ can be identified with Minkowski spacetime).
 %Then we can assume, by the continuity of $g$ and the local achronality of $S$, that there exists a connected neighbourhood $U \ni p$ in $M$ for which $U \cap S$ is an achronal spacelike submanifold in $(U, g|_U)$. But then $U \cap S$ is also acausal in $(U, g|_U)$ (see Lemma 42, p. 425, of \cite{oneill}), contradicting the clause (ii) in Definition \ref{def1.1}.
\qcd

A {\em (future, past) lightlike $C^k$ ($k \geq 0$) hypersurface} is simply a (future, past) lightlike set which is also a $C^k$ hypersurface in $M$. These have a structure very similar to that of {\em achronal boundaries}, i.e., sets of the form $\partial I^{\pm}(C)$ for some set $C \subseteq M$, as the next proposition shows.

Here and hereafter, we shall often state our results only in terms of future lightlike sets, in which case a time-dual version will be always be understood to hold.

\begin{proposition}
\label{prop1.3}
Assume that $S$ is a future lightlike $C^0$ hypersurface. Then:
\begin{itemize}
\item[i)] $S$ is the union of (the images of) maximal future-directed null geodesics without future endpoints in $S$ (when maximally extended in $S$, these will be called {\em null generators} of $S$).
\item[ii)] If $\alpha$ and $\beta$ are two distinct null generators, then either $\mbox{Im}\alpha \cap \mbox{Im} \beta =\emptyset$ or $\mbox{Im}\alpha \cap \mbox{Im} \beta = \{p\}$, $p\in S$ being their common past endpoint.
\item[iii)] If $S$ is achronal and $\gamma: [0,a) \rightarrow S$ is a null generator of $S$ starting at $p = \gamma(0)$, then there exists no conjugate points to $x_0$ along $\gamma$.
\item[iv)] If $S$ is closed, then all its null generators are future-inextendible (in $M$).
\end{itemize}
\end{proposition}
{\em Proof.} (i)
Let $p \in S$, and fix a neighbourhood $U \ni p$ in $M$ for which $U \cap S$ is achronal in $(U, g|_U)$. First note that since $S$ is a $C^0$ hypersurface, we can choose $U$ such that it is connected and $U\setminus S$ has two connected components. Indeed, we can assume, without loss of generality, that $U = U_{+} \cup (U \cap S) \cup U_{-}$, where $U_{\pm} := I^{\pm}(U \cap S, U)$ (for instance, by replacing $U$ by a chronological diamond contained in $U$ of two points at the past and the future of $U\cap S$). By (ii) in Defn. \ref{def1.1}, let $q \in U \cap S$, $q \neq p$, and $\tilde {\alpha} : [0,1] \rightarrow U$ a future-directed causal curve with $\tilde{\alpha}(0) = p$ and $\tilde{\alpha}(1) =q$. From the achronality of $U \cap S$, we can assume that $\tilde{\alpha}$ is a null geodesic without conjugate points to $p$ before $q$.

{\em Claim:} $\tilde{\alpha}[0,1] \subseteq U \cap S$.

Indeed, suppose not. Then, for some $t_0 \in (0,1)$, $\tilde{\alpha}(t_0) \in U \setminus (U\cap S)$. However, if $\tilde{\alpha}(t_0) \in U_{-}$, then for some $r \in U \cap S$, $p <_U  \tilde{\alpha}(t_0) \ll _U r$, and hence $p \ll _U r$, contradicting achronality. On the other hand, if $\tilde{\alpha}(t_0) \in U_{+}$, an analogous argument using $q$ instead $p$ also leads to a contradiction with achronality, thus proving the claim.

Let $\alpha : [0, a) \rightarrow M$ be the future-directed, future-inextendible (in $M$) null geodesic with $\alpha|_{[0,1]} = \tilde{\alpha}$ (so, in particular, $a>1$). If $\alpha[0,a) \subseteq S$ we are done, since the maximal extension of $\alpha$ in $S$ will then be a future-inextendible null generator passing through $p$. Otherwise, let
\[
s_0 = \sup \{ t \in [0,a) \, : \, \alpha [0,t] \subseteq S \}.
\]
By assumption $s_0 <a$, and by the previous Claim, $s_0 \geq 1$. Suppose by contradiction that $\alpha(s_0) \in S$. Then, for some neighbourhood $V \ni \alpha(s_0)$, $V \cap S$ is achronal in $(V,g|_{V})$, and reasoning as before, we can assume that there exists a number $\epsilon >0$ such that $s_0 + \epsilon <a$ and a future-directed null geodesic $\beta: [s_0,s_0 + \epsilon] \rightarrow V$ such that $\beta[s_0,s_0 + \epsilon] \subseteq V \cap S$ and $\beta(s_0) = \alpha(s_0)$. Now, by continuity, there exists a number $0< \delta < s_0$ for which $\alpha[s_0 - \delta, s_0] \subseteq V\cap S$. But then by achronality, we must have that $\beta \equiv \alpha |_{[s_0, s_0 + \epsilon]}$, and hence $\alpha[0,s_0 + \epsilon] \subseteq S$, in contradiction with the definition of $s_0$. Therefore, we conclude that $\alpha(s_0) \notin S$, and hence $\gamma := \alpha\mid_{[0,s_0)}$ has no future endpoint in $S$. So, its maximal extension in $S$ is a null generator passing through $p$.

(ii) Let $\alpha, \beta$ be null generators of $S$ with $\mbox{Im}\alpha \cap \mbox{Im} \beta \neq\emptyset$. By reparametrizing and restricting conveniently the domains of $\alpha, \beta$, we can assume that $\alpha:[0,a) \rightarrow M$, $\beta:[0,b) \rightarrow M$ with $\alpha(0)= \beta(0)=p$ and, say, $a\leq b$. Let
\[
T= \sup \{ t \in [0,a) \, : \, \alpha[0,t] \subseteq \beta[0,b) \}.
\]
If $T=a$ then $\mbox{Im}\alpha \subset \mbox{Im} \beta$, and the extensions of $\alpha$ and $\beta$ in $S$ coincide as null generators of $S$.

Suppose now that $T <a$. Assume by contradiction that $T>0$ and let $q = \alpha(T) \in S$. Again, for some neighbourhood $W \ni \alpha(T)$, $W \cap S$ is achronal in $(W,g|_{W})$, and reasoning just as in the previous part, we can assume that there exists a number $\epsilon >0$ such that $T + \epsilon <a$ and a future-directed null geodesic $\lambda: [T,T + \epsilon] \rightarrow W$ such that $\lambda[T,T + \epsilon] \subseteq W \cap S$ and $\lambda(T) = q$. Again by continuity, there exists a number $0< \delta < T$ for which $\alpha[T - \delta, T] \subseteq W\cap S$. Achronality now demands that $\lambda \equiv \alpha |_{[T, T + \epsilon]} = \beta |_{[T, T + \epsilon]}$, a contradiction. Therefore, $T=0$ and so $\mbox{Im}\alpha \cap \mbox{Im} \beta = \{p\}$. In this case, and taking into account local achronality, neither $\alpha$ nor $\beta$ can be extended as null generators to the past beyond $p$, which is then a common past endpoint.

(iii) is immediate.

(iv)
If $S$ is closed, a putative future endpoint of any null generator $\alpha:[0,a) \rightarrow M$ would be in $\overline{S} = S$, contradicting part (i).

\qcd

In the next section, it will be of importance to consider lightlike $C^0$ hypersurfaces whose null generators are {\em future-inextendible} (in $M$). Now, as Proposition \ref{Ex1.2} indicates, natural examples of future lightlike $C^0$ hypersurfaces are sets of the form $\partial I^{-}(C) \setminus \overline{C}$. However, in this case the null generators are not in general future-inextendible, having future endpoints on $\overline{C}$. On the other hand, $\partial I^{-}(C)$ is (if non-empty) an achronal, closed $C^0$ hypersurface, though in general not a lightlike set, as already observed. But if it is, then, being a closed lightlike $C^0$ hypersurface, it does have future-inextendible null generators (by the item (iv) of the previous proposition). So it becomes pertinent to investigate situations where this does occur. To simplify the nomenclature, we adopt the following definition.

\begin{definition}
\label{p-horizon}
A future lightlike set $A \subseteq M$ is said to have a {\em future p-horizon} if $\partial I^{-}(A)$ is a non-empty future lightlike closed $C^0$ hypersurface (which will then have future-inextendible null generators by Prop. \ref{prop1.3} (iv)). \footnote{Again, the time-dual concept of having a {\em past p-horizon} applies to past lightlike sets. The name ``p-horizon'' comes from the standard concept of {\em particle horizon} in Relativity (see, e.g., p. 128 of \cite{HE}).}
\end{definition}
An important first example of a future lightlike set having a future p-horizon is as follows.

\begin{proposition}
\label{prop1.4}
Let $S \subseteq M$ be a non-empty set, and let $\alpha: [0,a) \rightarrow M$ be a future-inextendible null $S$-ray. Then $\mbox{Im} \alpha$ has a future p-horizon.
\end{proposition}
{\em Proof.} The following proof is an adaptation of the first part of the proof of Theorem 4.1 in \cite{gallowaymaximum}. Write $A:= \mbox{Im} \alpha$. Since $A$ is achronal, $A \subseteq \partial I^{-}(A)$, so $\partial I^{-}(A)$ is a non-empty closed achronal $C^0$ hypersurface in $M$. So, it suffices to show that $\partial I^-(A)$ satisfies clause (ii) of Definition \ref{def1.1}.
Let $p \in \partial I^{-}(A)$, and let $U$ be a convex normal neighbourhood of $p$. Now, $(U,g|_U)$ is strongly causal, so given a pre-compact open neighbourhood $K$ of $p$, with $\overline{K}\subseteq U$, for each $t \in [0,a)$,   $\alpha|_{[t,a)}$ cannot remain in $K$ if it ever meets it. Thus, there exists a sequence $p_n = \alpha(t_n)$ with $t_n \nearrow a$, $p_n \notin K$. It follows that for each $x \in V = K \cap I^{-}(A)$, there exists a future-directed timelike curve from $x$ to a point on $A$ not in $V$. We apply now the (time-dual of) Lemma 3.19 of Ref. \cite{P} to conclude that $p$ is the past endpoint of a future-directed null geodesic segment contained in $\partial I^{-}(A)\cap U$.
\qcd

In order to generalize this example, we need the following concept.

\begin{definition}
\label{def1.2}
Let $A \subseteq M$ be any set. Its {\em future edge} is
\[
\epsilon_{+}A := \{ p \in \overline{A} \, : \, \exists U \ni p \mbox{ open such that } A \cap J^{+}(p,U) \subseteq \{p\} \}.
\]
(The {\em past edge} $\epsilon_{-}A$ is defined time-dually.)
\end{definition}

\begin{remark}
\label{remark1.1}
The future edge $\epsilon_{+}A$ is not necessarily closed. {\em To see this, consider the following example. Take $M = \mathbb{R}^3$, $g$ the Minkowski metric thereon, given by $ds^2_g = -dt^2 + dx^2 + dy^2$, and time-orientation such that $\partial / \partial t$ is future-directed. Let
\[
A = \{ (t,t,y) \, : \, 0 \leq t < 1, y \neq 0 \} \cup \{(t,t,0) \, : \, 0 \leq t \}.
\]
It is easy to see that $A$ is a future lightlike set, and that $\epsilon_{+}A = \{ (1,1,y) \, : \, y \neq 0 \}$, but $\overline{\epsilon_{+}A} = \{ (1,1,y) \, : \, y \in \mathbb{R} \}$. }
\end{remark}

\begin{proposition}
\label{prop1.5}
Let $A \subseteq M$ be any set.
\begin{itemize}
\item[i)] If $A$ is locally achronal, then $A$ is a future lightlike set iff $A \cap \epsilon_{+}A = \emptyset$.
\item[ii)] If $A$ is a future lightlike set, then $\partial I^{-} (A) \setminus  \overline{\epsilon_{+}A}$, if non-empty, is an achronal future lightlike $C^0$ hypersurface.
\end{itemize}
\end{proposition}
{\em Proof.} (i)
The ``only if'' part is immediate. For the ``if'' part, consider $p \in A$ and let $U \ni p$ be any neighbourhood for which $A \cap U$ is achronal in $(U,g|_{U})$. Since $p \notin \epsilon_{+}A$, there exists a point $q \neq p $ in $J^{+}(p, U) \cap A$. Since $p$ is arbitrary, $A$ is a future lightlike set.

(ii)
$S:= \partial I^{-} (A)\setminus  \overline{\epsilon_{+}A}$ is obviously achronal. Let $p \in S$. Since $p \notin \overline{\epsilon_{+}A}$, we can pick a neighbourhood $U \ni p$ for which $U \cap \overline{\epsilon_{+}A} = \emptyset $. In that case, $ S\cap U = \partial I^{-}(A) \cap U$, which is an achronal $C^0$ hypersurface. Moreover, since $p \notin \epsilon_{+}A$, there exists a point $q \neq p$ in $J^{+}(p,U) \cap A$, and the proof is complete.

\qcd

\begin{corollary}
\label{cor1.1}
If $A \subseteq M$ is a future lighlike set with $\epsilon_{+}A = \emptyset$ and $\partial I^{-} (A) \neq \emptyset$, then $A$ has a future p-horizon.
\end{corollary}

\qcd

\begin{corollary}
\label{cor1.2}
Assume that $(M,g)$ is strongly causal, and let $\eta: [0,a) \rightarrow M$ be a future-directed, future-inextendible null geodesic. If $\partial I^{-} (\eta)$ is non-empty, then $\eta$ has a future p-horizon.
\end{corollary}
{\em Proof.} Write $A:=\mbox{Im} \eta$. By Proposition \ref{Ex1.1} and the previous Corollary, we only need to show that  $\epsilon_{+}A = \emptyset$. But since $(M,g)$ is strongly causal, $A$ is closed, and hence $\epsilon_{+}A \subset A$. But by Proposition \ref{prop1.5}(i), $A \cap \epsilon_{+}A = \emptyset$, so the result follows.

\qcd

\begin{remark}
\label{remark1.2}
The converse of Corollary \ref{cor1.1} does not necessarily hold. {\em To see this, take again $M = \mathbb{R}^3$, $g$ the Minkowski metric thereon, given by $ds^2_g = -dt^2 + dx^2 + dy^2$. Let
\[
A = \{ (t,t,y) \, : \, 0 \leq t, y \neq 0 \} \cup \{(t,t,0) \, : \, 0 \leq t <1 \}.
\]
$A$ is a future lightlike set, and $\partial I^{-}(A) = \{ (t,t,y) \, : \, t,y \in \mathbb{R} \}$ is an achronal lightlike $C^{\infty}$ hypersurface, so $A$ has a future p-horizon. However, $(1,1,0) \in \epsilon_{+}A\neq\emptyset$. }
\end{remark}

\section{The geometry of null hypersurfaces}\label{2}

Throughout this section, $S$ shall denote a smooth null hypersurface (i.e., a codimension one embedded $C^{\infty}$ submanifold with everywhere degenerate induced metric) of $(M,g)$ \footnote{The results and concepts of this section are well known, and can be found, e.g., in \cite{beijing,kupeli}, which we closely follow. They have been included in this paper only for the sake of clarity, as well as to fix notation. Proposition \ref{main1}, however, is not in these references, although it is an almost immediate application of the basic tools we present here.}.

For each $p \in S$, the orthogonal complement $T_pS^{\perp}$ of the tangent space $T_pS$ of $S$ at $p$ within $T_pM$ is one-dimensional, and indeed $T_pS^{\perp} \subset T_pS$. This means that there is a canonical line bundle $TS^{\perp}$ associated with $S$ with fiber $T_pS^{\perp}$ at $p$. This line bundle clearly admits an everywhere non-zero smooth global section $K$, obtained, say, by fixing any future-directed timelike vector field $X:M \rightarrow TM$ and picking the value of $K$ at a given point $p \in S$ to be the unique null vector $K_p \in T_pS^{\perp}$ such that $g(K_p,X_p) = -1$. The future-directed null vector field $K:S \rightarrow TS$ is then simultaneously tangent and orthogonal to $S$, and is unique up to multiplication by a positive smooth function $f: S \rightarrow (0, + \infty)$.

The following fact is fundamental (see, e.g., Corollaries 14 and 15 of \cite{kupeli} for a proof).

\begin{proposition}
\label{prop2.1}
If $S$ is a smooth null hypersurface in the spacetime $(M,g)$ and $K:S \rightarrow TS$ is a future-directed null vector field on $S$, then $\nabla_K K = \lambda K$ for some $\lambda \in C^{\infty}(S)$. In particular, the integral curves of $K$ are null pre-geodesics of $(M,g)$.
\end{proposition}

The null geodesics reparametrizing integral curves of $K$ are called the {\em null (geodesic) generators} of $S$. They are intrinsic to $S$.

To study the ``shape'' of the null hypersurface $S$, we study how the null vector field $K$ varies along $S$. To this end we shall introduce a ``mod $K$'' version of the shape operator (Weingarten map) and associated second fundamental form.

Now, the bundle $TS/TS^{\perp} := \cup _{p \in S} T_pS/T_pS^{\perp}$ is a smooth rank $n-1$ vector bundle over $S$. It admits a natural positive-definite fiber metric $h$, given by
\[
h_p(\overline{v},\overline{w}) := g_p(v,w),
\]
for each $v,w \in T_pS$ and each $p \in S$, where the bar over the vector denotes its equivalence class mod $T_pS^{\perp}$. It is very easy to check that it is well-defined.

The {\em null Weingarten map $b = b_K$ (of $S$ with respect to $K$)} for each point $p \in S$ is the linear map $b: \overline{v} \in T_pS/T_pS^{\perp} \mapsto \overline{ \nabla _v K} \in T_pS/T_pS^{\perp}$\footnote{For ease of notation, we will often omit an explicit reference to the point $p \in M$.}. This map is well-defined, because given equivalent $v,v' \in T_pS$, we have $v' = v + a K_p$ for some number $a$, and then
\[
\overline{\nabla _{v'} K} = \overline{ \nabla_{v}K + a \nabla_{K_p} K} = \overline{ \nabla_{v}K},
\]
where we have used Proposition \ref{prop2.1} to obtain the last equality.

\begin{proposition}
\label{prop2.2}
If $S$ is a smooth null hypersurface in the spacetime $(M,g)$ and $K:S \rightarrow TS$ is a future-directed null vector field on $S$ with associated Weingarten map $b$, then
\begin{itemize}
\item[i)] $b$ is self-adjoint with respect to $h$, i.e., $h(b(\overline{v}),\overline{w}) = h(b(\overline{w}),\overline{v})$, for all $\overline{v}, \overline{w} \in T_pS/T_pS^{\perp}$;
\item[ii)] if $f: S \rightarrow (0, +\infty)$ is smooth, so that $\tilde{K} = f K$ is another future-directed null vector field on $S$, then $b_{\tilde{K}} = f b_K$.
\end{itemize}
\end{proposition}
{\em Proof.} (i)
Extend $v,w \in T_pS$ to smooth vector fields $V,W$ on $M$ tangent to $S$ near $p$, respectively. Since $g(K,W)\equiv g(K,V)\equiv 0$ on $S$ near $p$,
\[
\begin{array}{l}
Vg(K,W)(p)=g(\nabla_V K,W)(p)+g(K,\nabla_V W)(p)=0, \\ Wg(K,V)(p)=g(\nabla_W K,V)(p)+g(K,\nabla_W V)(p)=0.
\end{array}
\]
Therefore,
\begin{eqnarray}
h(b(\overline{v}), \overline{w}) &=& g(\nabla_V K,W)(p) = -g(K,\nabla_V W)(p) =  -g(K, \nabla_W V)(p) - g(K,[V,W])(p) \nonumber \\
&=& -g(K, \nabla_W V)(p) = g(\nabla_W K,V)(p) = h(b(\overline{w}),\overline{v}). \nonumber
\end{eqnarray}
(ii)
For each $v\in T_pS$,
\[
b_{\tilde{K}}(\overline{v}) = \overline{ \nabla _v \tilde{K}} = \overline{v(f)K + f \nabla_v K}=\overline{f \nabla_v K}\equiv f b_{K}(\overline{v}).
\]

\qcd

The {\em null second fundamental form $B = B_K$ (of $S$ with respect to $K$)} is the bilinear form associated with $b$ via $h$, i.e., $B:T_pS/T_pS^{\perp} \times T_pS/T_pS^{\perp} \rightarrow \mathbb{R}$ is given by
\[
B(\overline{v}, \overline{w}) = h(b(\overline{v}),\overline{w}) = g(\nabla_{v}K,w),\qquad\hbox{for all $\overline{v},\overline{w} \in T_pS/T_pS^{\perp}$.}
\]
Since $b$ is self-adjoint, $B$ is symmetric. We say that $S$ is {\em totally geodesic} iff $B \equiv 0$ (equivalently, $b \equiv 0$) everywhere on $S$. One can show this has the usual geometric meaning: if $S$ is totally geodesic, then given any geodesic $\gamma:[0,a) \rightarrow M$ starting at some $p \in S$ with $\gamma '(0) \in T_pS$, there exists $0< \epsilon < a$ for which $\gamma[0, \epsilon) \subseteq S$.

The {\em null mean curvature}, or {\em null expansion scalar} ({\em of $S$ with respect to $K$}) is $\theta = \theta_K = \mbox{tr} b$. If we pick any $g$-orthonormal set $\{e_1, \ldots, e_{n-1} \} \subset T_pS$ of spacelike vectors, then $\{\overline{e_1}, \ldots, \overline{e_{n-1}} \}$ is an $h$-orthornomal basis for $T_pS/T_pS^{\perp}$. Hence
\begin{eqnarray}
\theta(p) = \mbox{tr} b &=& \sum_{i=1}^{n-1} h_p(b(\overline{e_i}),\overline{e_i}) = \sum_{i=1}^{n-1} g_p(\nabla _{e_i}K, e_i) \nonumber \\
&=& \mbox{div}_\Sigma K, \nonumber
\end{eqnarray}
where $\Sigma$ is any smooth spacelike codimension 2 submanifold $\Sigma \subset S$ such that $\mbox{span}\{e_1, \ldots, e_{n-1} \} = T_p\Sigma$. Therefore, $\theta$ measures the expansion of the null generators of $S$ towards the future. Moreover, it follows from Proposition \ref{prop2.2}(ii) that null mean curvature inequalities such as $\theta \geq 0$, $\theta <0$, etc., are invariant under positive rescaling of $K$. For example, according to our conventions, in Minkowski spacetime a future null cone $S = \partial I^{+}(p) \setminus \{p \}$ (resp. past null cone $S = \partial I^{-}(p) \setminus \{p \}$) always has positive null mean curvature $\theta >0$ (resp. negative null mean curvature $\theta <0$).

We now describe how the Weingarten map propagates along the null geodesic generators. Let $\eta: I \subseteq \mathbb{R} \rightarrow S$ be a future-directed, affinely parametrized null geodesic generator of $S$. For each $s \in I$, denote by $b(s)$ the Weingarten map based at $\eta(s)$. Then we can show that $b(s)$ satisfies a Ricatti-type equation (cf. Proposition 3.3., p. 35 of Ref. \cite{beijing}).

\begin{proposition}
\label{ricatti}
The one-parameter family of Weingarten maps $s \in I \mapsto b(s)$ obeys the equation
\begin{equation}
\label{ricattiequation}
b' + b^2 + R = 0,
\end{equation}
where $b'$ is the covariant derivative, $b^2 = b \circ b$ and $R:T_{\eta(s)}S/T_{\eta(s)}S^{\perp} \rightarrow T_{\eta(s)}S/T_{\eta(s)}S^{\perp}$ is the {\em curvature endomorphism}, defined by $R(\overline{v}) = \overline{R(v, \eta '(s))\eta '(s)}$.
\end{proposition}

By taking the trace of Eq. (\ref{ricattiequation}), we obtain the following formula for the derivative of the null mean curvature $\theta_{\eta}(s) := \theta (\eta(s))$ along $\eta$:
\begin{equation}
\label{raychaudhuri}
\theta_{\eta} ' + \frac{1}{n-1} \theta_{\eta}^2 + \sigma ^2 + Ric(\eta',\eta') = 0,
\end{equation}
where $\sigma:= (\mbox{tr} \hat{b} ^2)^{1/2}$ is the {\em shear scalar}, being $\hat{b}:= b - \frac{1}{n-1} \theta \cdot Id$ is the trace-free part of the Weingarten map, and $Ric$ is the Ricci tensor of $(M,g)$. Eq. (\ref{raychaudhuri}) is the Raychaudhuri equation for (twist-free) null geodesics.

With these tools in place, we can prove the following.

\begin{proposition}
\label{main1}
Suppose that $(M,g)$ satisfies the null convergence condition, i.e., $Ric(v,v) \geq 0$ $\forall v \in TM$ null. Let $S\subset M$ be a null hypersurface, and suppose that there exists a smooth submanifold $\Sigma$ of $S$ such that:
\begin{itemize}
\item[1)] $\Sigma$ is a spacelike, acausal codimension 2 {\em maximal} (i.e., zero mean curvature) submanifold of $(M,g)$ (hence codimension 1 in $S$);
\item[2)] Every null generator of $S$ intersects $\Sigma$ orthogonally.
\end{itemize}
Then
\begin{itemize}
\item[i)] $S$ is diffeomorphic to $\Sigma \times \mathbb{R}$, $S_{\pm} := (S \cap J^{\pm}(\Sigma)) \setminus \Sigma$ are disjoint smooth null hypersurfaces, and $S = S_{+} \cup \Sigma \cup S_{-}$.
\item[ii)] $S_{+}$ has null mean curvature $\theta \leq 0$.
\end{itemize}
\end{proposition}
{\em Proof.} (i) From condition (2) and the acausality of $\Sigma$, every null generator of $S$ must intersect $\Sigma$ precisely at one parameter value. By Lemma 15 of \cite{kupeli}, $S$ is diffeomorphic to $\Sigma \times \mathbb{R}$. In particular, $\Sigma$ is closed in $S$, so $S \setminus \Sigma$ is an open submanifold of $S$. From the acausality of $\Sigma$, $S_{+}\cap S_{-} = \emptyset$ . Finally, condition (2) yields $S \setminus \Sigma \subseteq S_{+}\cup S_{-}$.

(ii)
Let $K:S \rightarrow TS$ be a tangent, smooth, future-directed, null vector field on $S$. Since the integral curves of $K$ differ from the null generators only by a reparametrization, $K|_{\Sigma}$ is orthogonal to $\Sigma$. Let $p \in S_{+}$ and let $\eta:(a,b) \rightarrow S$ be the (maximally extended) null generator of $S$ passing through $p$. We can assume, by a suitable choice of the affine parametrization, that $a<0<b$ and $\eta(0) \in \Sigma$, and thus, for some $t_0 \in (0,b)$, $\eta(t_0) = p$. Since $\Sigma$ has zero mean curvature and $K|_{\Sigma}$ is normal to $\Sigma$, $\theta(\eta(0))=0$. Thus, Eq. (\ref{raychaudhuri}) together with the null convergence condition imply that $\theta(\eta(s)) \leq 0$ for all $s \in [0,b)$, and in particular $\theta(p)=\theta(\eta(t_0))\leq 0$.

\qcd

\section{Comparison theory for lightlike $C^0$ hypersurfaces}\label{3}

We start with the following definition, due to Galloway (cf. Definition 3.2 of \cite{gallowaymaximum}).
\begin{definition}
\label{support}
Let $S \subset M$ be a future lightlike $C^0$ hypersurface. We say that $S$ has {\em null mean curvature $\theta \geq 0$ in the support sense} provided that for each $p \in S$ and for each number $\epsilon >0$ there exists a smooth (at least $C^2$) null hypersurface $S_{p,\epsilon} \ni p$ such that
\begin{itemize}
\item[(i)] for some neighbourhood $U \ni p$ for which $S \cap U$ is achronal in $(U, g|_{U})$, we have $S \cap U \subseteq J^{+}(S_{p, \epsilon} \cap U,U)$, and
\item[ii)] the null mean curvature $\theta_{p, \epsilon}$ of $S_{p, \epsilon}$ (wrt to the unique future-directed null vector field on $S_{p, \epsilon}$ having unit norm wrt a fixed background Riemannian metric) satisfies $\theta_{p, \epsilon}(p) \geq - \epsilon$.
\end{itemize}
\end{definition}
(If $S \subset M$ is a {\em past} lightlike $C^0$ hypersurface, then it can be defined in a time-dual fashion what it means to say that $S$ has {\em null mean curvature $\theta \leq 0$ in the support sense}.)

The next two theorems are a suitable transcription of Lemma 4.2 and Theorem 3.4 in \cite{gallowaymaximum}, respectively.

\begin{theorem}
\label{comparison1}
Suppose that $(M,g)$ satisfies the null convergence condition, and let $S \subset M$ be an achronal future lightlike $C^0$ hypersurface whose null generators are all future-complete. Then $S$ has null mean curvature $\theta \geq 0$ in the support sense, with null second fundamental forms locally bounded from below\footnote{This is a technical condition which arises in the statement of the maximum principle for $C^0$ null hypersurfaces (Theorem \ref{comparison2}), and can be stated as follows (cf. \cite{gallowaymaximum}): let $B_{p,\epsilon}$ denote the null second fundamental form of $S_{p,\epsilon}$ at $p$. We say that the collection of null second fundamental forms $\{B_{p,\epsilon}: p\in S, \epsilon>0\}$ is {\em locally bounded from below} provided that for all $p\in S$ there is a neighborhood $W$ of $p$ in $S$ and a constant $k>0$ such that $B_{q,\epsilon}\geq -k h_{q,\epsilon}$ for all $q\in W$ and $\epsilon>0$,
where $h_{q,\epsilon}$ is the Riemannian metric on $T_q S_{q,\epsilon}/K_{q,\epsilon}$.}.
\end{theorem}

\begin{theorem}[Maximum Principle for $C^0$ null hypersurfaces]
\label{comparison2}
Let $S_1 \subset M$ be a future lightlike $C^0$ hypersurface and $S_2 \subset M$ be a past lightlike $C^0$ hypersurface. Assume that for some $p \in M$,
\begin{itemize}
\item[(i)] $p \in S_1 \cap S_2$ and there exists a neighbourhood $U \ni p$ such that $S_i \cap U$ ($i=1,2$) is achronal in $(U,g|_{U})$ and $S_2 \cup U \subseteq J^{+}(S_1 \cup U,U)$,
\item[(ii)] $S_1$ (resp. $S_2$) has null mean curvature $\theta_1 \geq 0$ (resp. $\theta_2 \leq 0$) in the support sense, with null second fundamental forms locally bounded from below.
%\footnote{Actually, an additional technical hypothesis on the null fundamental forms of the support surfaces for $S_1$ is required in Theorem \ref{comparison2}. This, however, will not be important for us, since Lemma 4.2 in \cite{gallowaymaximum} in fact guarantees that this technical assumption holds in our context. Thus, for simplicity I have omitted it in the statements of both Theorem \ref{comparison1} and Theorem \ref{comparison2}.}.
\end{itemize}
Then there exists a neighbourhood $O \ni p$ for which $S:=S_1 \cap O = S_2 \cap O$. Moreover, $S$ is a smooth null hypersurface with null mean curvature $\theta \equiv 0$.
\end{theorem}

The following result is a simple consequence of the Theorem 3.2.31 of \cite{treude}.

\begin{proposition}
\label{nullcutlocus}
Suppose $(M,g)$ is globally hyperbolic and let $\Sigma \subset M$ be an acausal, future causally complete\footnote{Recall that a subset $C \subseteq M$ is {\em future causally complete} (FCC) if for all $p \in J^{+}(C)$, the set $J^{-}(p) \cap C$ has compact closure in $C$. It is easy to check that every FCC subset of $M$ is closed.} submanifold. Then $S = \partial J^{+}(\Sigma)\setminus (\Sigma \cup \mbox{Cut}(\Sigma))$ is a smooth null hypersurface, where Cut($\Sigma$) is the set of null cut points to $\Sigma$ along future-directed null geodesics starting at and normal to $\Sigma$.
\end{proposition}

\begin{lemma}
\label{lemmamain2}
Assume that $(M,g)$ is globally hyperbolic and let $C \subseteq M$ be a future causally complete set. Then $J^{+}(C)$ is a closed set.
\end{lemma}
{\em Proof.} Let $(x_n) \subseteq J^{+}(C)$ be a sequence converging to a point $x \in M$. Pick any $x_{+} \in I^{+}(x)$, so we can assume without loss of generality that $x_n \in I^{-}(x_{+}) \cap J^{+}(C)$. Now, the latter set is easily seen to be contained in the set $D:= J^{-}(x_{+}) \cap J^{+}(K)$, where $K := J^{-}(x_{+}) \cap C$, which is compact because $C$ is $C$ is future causally complete. Since $(M,g)$ is globally hyperbolic, $D$ is also compact, so $x \in J^{+}(C)$ and the result follows.
\qcd

\begin{corollary}
\label{comparison3}
In the conditions and notation of Proposition \ref{nullcutlocus}, if $(M,g)$ satisfies the null convergence condition and $\Sigma$ is in addition a codimension 2 maximal submanifold, then $S=\partial J^+(\Sigma)\setminus (\Sigma\cup \mbox{Cut}(\Sigma))$ has null mean curvature $\theta \leq 0$.
\end{corollary}
{\em Proof.} Extending the null generators of $S$ a little to the past, we can easily obtain a null hypersurface which will satisfy the conditions of Proposition \ref{main1} (we need $J^+(C)$ closed to ensure (2) in that proposition), and the result follows.
\qcd

Our first main result is the following theorem.

\begin{theorem}
\label{main2}
Suppose $(M,g)$ is globally hyperbolic, future null geodesically complete and satisfies the null convergence condition. Let $\Sigma \subset M$ be an acausal, future causally complete, maximal codimension 2 spacelike submanifold. Suppose furthermore that $\partial I^{+}(\Sigma)\setminus \Sigma$ contains a connected future lightlike set $A$ with a future p-horizon. Then, the connected component of $\partial I^{+}(\Sigma) \setminus \Sigma$ containing $A$ is a smooth totally geodesic null hypersurface with future-complete null geodesic generators.
\end{theorem}
{\em Proof.} We adapt herein the proof of Theorem 7.1 of \cite{EGP}. Note first that $A$ is an achronal set (because $A\subset \partial I^+(\Sigma)$, and $\partial I^+(\Sigma)$ is achronal), and hence $A \subseteq \partial I^{-}(A)$. Since $A$ has a future p-horizon, the connected component $S_1$ of $\partial I^{-}(A)$ containing $A$ is a non-empty, achronal, closed, future lightlike $C^0$ hypersurface with future-inextendible null generators, which are therefore future-complete. By Theorem \ref{comparison1}, $S_1$ has null mean curvature $\theta_1 \geq 0$ in the support sense, with null second fundamental forms locally bounded from below.

We claim that $A \subseteq \partial I^{+}(\Sigma)\setminus (\Sigma \cup \mbox{Cut}(\Sigma))$. Indeed, let $p \in A$. Since $p \in  \partial I^{+}(\Sigma)\setminus \Sigma$, and $\Sigma$ is future causally complete, $J^+(\Sigma)$ is closed (by Lemma \ref{lemmamain2}), and the null geodesic generator $\eta$ of $\partial I^{+}(\Sigma)$ passing through $p$ has a past endpoint on $\Sigma$, having no focal points to $\Sigma$ before $p$. Since $A$ is future lighlike, $\eta$ can be extended a little further beyond $p$ without focal points, and therefore $p \notin \mbox{Cut}(\Sigma)$.

Let $S_2$ be the connected component of $\partial I^{+}(\Sigma)\setminus (\Sigma \cup \mbox{Cut}(\Sigma))$ containing $A$. By Proposition \ref{nullcutlocus} and Corollary \ref{comparison3}, $S_2$ is a smooth null hypersurface with null mean curvature $\theta_2 \leq 0$.

Next, we claim that $S_2 \subset S_1$. First, note that $A\subset S_1 \cap S_2\neq\emptyset$. So, it suffices to prove that $S_1\cap S_2$ is open and closed in the connected set $S_2$. In order to prove that $S_1\cap S_2$ is closed in $S_2$, let $(x_n) \subseteq S_1 \cap S_2$ be a sequence converging to a point $x \in S_2$. Since $S_1$ is closed, $x \in S_1$. Hence $S_1 \cap S_2$ is closed in $S_2$. In order to prove that it is open, let $x \in S_1 \cap S_2$. Since $S_1$ is an achronal $C^0$-hypersurface, it is edgeless (cf., e.g., Proposition 25, Ch. 14 of \cite{oneill}). Hence, there exists an open set $U \ni x$ such that every future-directed timelike curve contained in $U$ and going from $I^{-}(x,U)$ to $I^{+}(x,U)$ must intersect $S_1$. Fix $z_{\pm} \in I^{\pm}(x,U)$ and let $V:= I^{-}(z_{+},U) \cap I^{+}(z_{-},U)$. Clearly, every point $y \in V \setminus S_1$ must be either in $I^{+}(S_1 \cap V,V)$ or in $I^{-}(S_1 \cap V,V)$ (but not in both, because $S_1$ is achronal). Moreover, $I^-(S_1\cap V,V)\cap S_2=\emptyset$, since otherwise there would exist a past-directed timelike curve from $A \subseteq\partial I^{+}(\Sigma)\setminus \Sigma $ to $S_2 \subset \partial I^{+}(\Sigma)\setminus \Sigma$, violating the achronality of $\partial I^{+}(\Sigma)$. Therefore, $(V\cap S_2)\setminus S_1=(V\setminus S_1)\cap S_2\subset I^{+}(S_1\cap V,V)\cap S_2$, and thus,
$S_2 \cap V\subseteq I^{+}(S_1 \cap V,V)\cup S_1 \subseteq J^{+}(S_1 \cap V,V)$. By the maximum principle for $C^0$ null hypersurfaces (Theorem \ref{comparison2}), there exists a neighbourhood $O \ni x$ for which $S_1 \cap O = S_2 \cap O$, this intersection being a smooth null hypersurface with null mean curvature $\theta \equiv 0$. Therefore, $S_1 \cap S_2$ is open in $S_1$ and in $S_2$, and thus, $S_1 \cap S_2 = S_2$, that is, $S_2 \subset S_1$. In particular, we have shown that $S_2$ is a smooth null hypersurface with future-inextendible null generators and $\theta\equiv 0$.

By the Eq. (\ref{raychaudhuri}) and the null convergence condition, along any generator $\eta$ of $S_2$ we must have $\sigma^2 = Ric(\eta', \eta') \equiv 0$, and hence $b\equiv 0$. In particular, $S_2$ is totally geodesic. Therefore, if $S_2$ coincides with the connected component $\tilde{S}_2$ of $\partial I^+(\Sigma)\setminus \Sigma$ containing $A$, then it must be a smooth null hypersurface with future-complete null geodesic generators and zero null mean curvature, and we are done. So, let us see that $S_2$ coincides with $\tilde{S}_2$. Note that $S_2\subset \tilde{S}_2$ and it is open in $\tilde{S}_2$ (because it is a submanifold of the same dimension). By contradiction, let $(y_n)\subset S_2$ be a sequence such that $y_n\rightarrow y\in \tilde{S}_2$, $y\not\in S_2$. Then, $y\in \overline{S}_2\cap \tilde{S}_2$, and so, $y\in\overline{S_2}\cap \mbox{Cut}(\Sigma)$. Since $S_1$ is closed and $S_2\subset S_1$, we have that $\overline{S}_2\subset S_1$. Hence, $y\in S_1\cap \mbox{Cut}(\Sigma)$, and thus, $y\in \partial I^-(A)\cap\mbox{Cut}(\Sigma)$.
Let $\alpha$ be a future-directed null generator of $\partial I^+(\Sigma)$ from $\Sigma$ to $y$, and let $\beta$ be a future-directed generator of $\partial I^-(A)$ starting at $y$. If $\beta$ is not the continuation of $\alpha$, then $\beta$ would enter in $I^+(\Sigma)$. But $\beta\subset \partial I^-(A)$, hence $I^+(\Sigma)\cap \partial I^-(A)\neq\emptyset$, and thus, $I^+(\Sigma)\cap I^-(A)\neq \emptyset$, in contradiction with the achronality of $\partial I^+(\Sigma)$.
%Now, there are two possibilities: either $y$ is a focal point, or it is the intersection with another generator $\beta$ of $\partial I^-(A)$. Since $\gamma$ is temporal, there exists $q\in\Sigma$ such that $q\ll y'$, thus $q\ll r\in A$. Therefore, $\alpha=\beta$, and thus, there are no focal points.
%In conclusion, $S_2$ coincides with the connected component $\tilde{S}_2$ of $\partial I^{+}(\Sigma) \setminus \Sigma$ containing $A$, which is thus a smooth null hypersurface with future-complete null geodesic generators and zero null mean curvature.
\qcd

%In conclusion, the future-directed null geodesic generators of $S_2$ cannot have future endpoints in $\mbox{Cut}(\Sigma)$, for they must coincide with the null generators of $S_1$ at all points of $S_2$. In other words, $\overline{S_2} \cap \mbox{Cut}(\Sigma) = \emptyset$.

\section{A rigidity result for generalized plane waves}\label{4}

%\footnote{The introduction to this section is essentially extracted from \cite{FS}.}
Throughout this section we shall assume that $(M,g)$ is a {\em generalized plane wave}, i.e, $M=M_0 \times \mathbb{R}^2$ and
\begin{equation}\label{pfw}
g(.,.) = g_0(.,.) + 2dudv + H(x,u)du^2,
\end{equation}
where $g_0$ is a smooth Riemannian metric on the ($n-1$)-dimensional manifold $M_0$, the variables $(v,u)$ are the standard coordinates of $\mathbb{R}^2$, and $H:M_0 \times \mathbb{R} \rightarrow \mathbb{R}$ is a smooth real function.

It is straightforward to check that the vector field $\partial_v$ is absolutely parallel (i.e.,
covariantly constant) and null, and the time-orientation will
be chosen which makes it past-directed. Thus, for any future-directed
causal curve $z(s)=(x(s),v(s),u(s))$,
\[
\dot u(s) = g(\dot z(s), \partial_v) \geq 0,
\]
and the inequality is strict if $z(s)$ is timelike. Since  grad$\,u
=\partial_v$, the coordinate $u: M \rightarrow {\mathbb R}$ plays the role of
a ``quasi-time'' function, i.e., its gradient is everywhere causal
and any causal segment $\gamma$ with $u\circ \gamma$ constant
(necessarily a null pregeodesic without conjugate points) is
injective. In particular, every generalized plane wave is causal. The
hypersurfaces $u\equiv\hbox{constant}$ are degenerate, with the
kernel of the metric spanned by $\partial_{v}$. The hypersurfaces
(non-degenerate $(n-1)$-submanifolds) of these degenerate
hypersurfaces which are transverse to $\partial_v$, must be
isometric to open subsets of $M_0$.

According to  Ehlers and Kundt \cite{EK} (see also \cite{Bi}) a
vacuum spacetime is a {\em plane-fronted gravitational wave} if it
contains a shearfree geodesic null vector field $K$, and
admits ``plane waves'' --spacelike (two-)surfaces  orthogonal to
$K$. The best known subclass of these waves are the
(gravitational) ``plane-fronted waves with parallel rays'' or
``pp-waves'', which are characterized by the condition that
 $K$ is covariantly constant, $\nabla K=0$.
Ehlers and Kundt gave several characterizations of these waves in
coordinates, and they obtained that ``at least locally'' the
metric can be written as in (\ref{pfw'}). Nowadays,
pp-wave means any spacetime which admits a covariantly constant
null vector field \cite[p. 383]{SKMHH}. Even though, in
general, their fronts may be ``non-plane'', this happens in the
most relevant cases (four dimensional spacetimes which are either
vacuum, or solutions to Einstein-Maxwell equations, or pure
radiations fields).
%Does it mean that the universal cover of any geodesically complete
%vacuum spacetime with a parallel lightlike vector field must admit
%the expression (\ref{pfw'}) with $\Delta_xH(x,u)$ vanishing? If
%true, we might rewrite the theorem in the introduction in the
%following nice form: {\em any geodesically complete vacuum
%spacetime with a parallel lightlike vector field must be a plane
%wave or a quotient of it.}}.
In what follows, we will use the term
{\em pp-wave} to denote the classical spacetimes
\begin{equation}\label{pfw'}
\begin{array}{c}
M= {\mathbb R}^{n-1}\times{\mathbb R}^2 \\
g_{pp}= dx_1^2+\cdots+dx_{n-1}^2+2\ du\ dv + H(x,u)\ du^2.
\end{array}
\end{equation} The pp-wave is vacuum (i.e., Ricci-flat) if and only if the ``spatial''
(transverse) Laplacian $\Delta_0H(x,u)$ vanishes\footnote{We shall denote quantities defined wrt the metric $g_0$ by mean of a superscript or a subscript ``$0$''.}.

Fixing some local coordinates $x_1, \dots, x_{n-1}$ for the Riemannian
part $M_0$ in (\ref{pfw}), it is straightforward to compute the
Christoffel symbols of $g$ and, thus,
to relate the Levi-Civita connections $\nabla, \nabla^0$ for $M$
and $M_0$, respectively (see \cite{CFSgrg}). We remark the following
facts:
\begin{itemize}
\item $M_0$ is totally geodesic, i.e., $\nabla_{\partial i}
\partial_j = \nabla^0_{\partial_{i}} \partial_j $, $i,j=1, \dots, n-1$.
\item The non-zero curvature coefficients are:
\[
R^i_{jkl} = (R_{g_0})^i_{jkl}, \quad R^i_{uuk} = \frac{1}{2}(g_0)^{il}(Hess_0H)_{lk}, \quad R^v_{iuj} = -\frac{1}{2}(Hess_0H)_{ij},
\]
where we have used $i,j,k,l, \ldots$ for the spatial coordinates, and $Hess_0H$ denotes the Hessian of $H = H( . , u)$ with respect to the metric $g_0$.

\item The Ricci tensor
$Ric$ of $(M,g)$ and ${\rm
Ric}^{0}$ of $(M_0,g_0)$ satisfy
\[
Ric = \sum_{i,j=1}^{n-1} R^{0}_{i j} d x_i \otimes d x_j
-\frac{1}{2}\Delta_{x}H du \otimes du .
\]
Thus, $Ric$ is zero if and only if both the Riemannian
Ricci tensor Ric$^{0}$ and the transverse Laplacian $\Delta_{0}H$
vanish.
\end{itemize}

From the direct computation of Christoffel symbols of a generalized plane wave, it is straightforward to write the geodesic equations
in local coordinates. Remarkably, the three geodesic equations for
a curve $z(s)= (x(s), v(s), u(s))$, $s\in ]a,b[$, can be solved in
the following three steps \cite[Proposition 3.1]{CFSgrg}:
\begin{enumerate}
\item[(a)] $u(s)$ is any affine function, $u(s) = u_0 + s \Delta
u$, for some $\Delta u\in {\mathbb R}$.

\item[(b)] Then $x = x(s)$ is a solution on $M_0$ of
\[
D_s\dot x = - {\rm grad}_x V_{\Delta}(x(s),s) \quad \mbox{for all
$s \in \ ]a,b[$,}
\]
where $D_s$ denotes the covariant derivative and $V_{\Delta}$ is
defined as:
\[
V_{\Delta}(x,s) = -\ \frac{(\Delta u)^2}{2}\ H(x, u_0 + s \Delta
u);
\]

\item[(c)] Finally, with a fixed $v_{0}$ and an $s_0\in ]a,b[$,
$v(s)$ can be computed from:
\[
v(s) = v_0 + \frac{1}{2 \Delta u} \int_{s_0}^s \left( E_z -
g_0(\dot x(\sigma), \dot x(\sigma)) + 2
V_{\Delta}(x(\sigma), \sigma)\right) d\sigma.
\]
where $E_z=g(\dot z(s), \dot z(s))$ is a
constant (if $\Delta u = 0$ then $v = v(s)$ is also affine).
\end{enumerate}

%..........
%
%Recall that a pp-wave spacetime consists of $\R^{4}$ endowed with
%metric
%\begin{equation}\label{metric-pp}
%g_{pp}=2\,du\,dv+H(x_{2},x_{3},u)du^{2}+dx_{2}^{2}+dx_{3}^{2},\qquad
%(u,v,x_{2},x_{3})\in\R^{4}.
%\end{equation}
%In particular, if we denote $x\equiv (x_{2},x_{3})$:
%\begin{itemize}
%\item[$(H_{1})$] $(\R^{4},g_{pp})$ is empty iff
%$\Delta_{x}H(x,u)=0$ for all $x,u$; \item[$(H_{2})$]
%$(\R^{4},g_{pp})$ is complete iff the trajectories $x(s)$
%solutions of $D_{s}\dot{x}=\frac{1}{2}\nabla_{x}H(x,s)$ are
%complete.
%\end{itemize}
%On the other hand, a gravitational plane wave is a pp-wave
%spacetime with
%\begin{equation}\label{z}
%H(x_{2},x_{3},u)=f(u)(x_{2}^{2}-x_{3}^{2})+2g(u)x_{2}x_{3}.
%\end{equation}
%Therefore, the problem reduces to prove that under conditions
%$(H_{1})$, $(H_{2})$ any pp-wave $(\R^{4},g_{pp})$ can be
%expressed with $H$ as in (\ref{z}).

Our main result in this section is the following.

\begin{theorem}
\label{main3}
Assume that $(M,g)$ is a globally hyperbolic, future null geodesically complete generalized plane wave obeying the null convergence condition. Fix $(v_0, u_0) \in \mathbb{R}^2$, and let
\[
\Sigma = \Sigma_{v_0,u_0} = \{(x,v_0,u_0) \in M \, | \, x \in M_0 \}.
\]
Suppose
\begin{itemize}
\item[i)] $\Sigma$ is entirely contained in the causal future of some Cauchy hypersurface \footnote{The only part of the following proof in which this hypothesis is used is when we prove that $\Sigma$ is FCC. Thus, we could have assumed, alternatively, that this is the case.}$V$, and
\item[ii)] there exists a future-directed null $\Sigma$-ray $\eta:[0,\infty) \rightarrow M$ starting at $\Sigma$ and not contained in the hypersurface $u = u_0$.
\end{itemize}
Then, $\partial I^{+}(\Sigma) \setminus \Sigma$ is the disjoint union of two connected smooth totally geodesic null hypersurfaces $S_{\pm}$ with future-complete null geodesic generators, where
\begin{itemize}
\item[a)] $S_{-} = (J^{+}(\Sigma) \setminus \Sigma) \cap \{u =u_0\}$ and
\item[b)] $S_{+}$ is the connected component of $\partial J^{+}(\Sigma)\setminus \Sigma$ containing $\eta(0,\infty)$.
\end{itemize}
Moreover, $\S= \partial J^{+}(\Sigma) = S_{+}\cup \Sigma \cup S_{-}$ is a {\em future Cauchy hypersurface} in $(M,g)$, i.e. an achronal set $\S$ for which $J^{+}(\S)= D^{+}(\S)$, and Hess$_0 H$ vanishes on the whole region $u>u_0$.

\end{theorem}
{\em Proof.} Note that $\Sigma$ is clearly a smooth codimension 2 closed spacelike submanifold of $M$ contained in the null achronal hypersurface $\S_0:=\{u=u_0\}$. We already know that $\Sigma$ is totally geodesic, hence maximal. Similarly, a straightforward calculation also shows that $\S_0$, and hence $S_{-}\subset\S_0$, is a smooth totally geodesic null hypersurface with future-complete null generators of the form $\gamma(v)=(v,u_0,x_0)$.

Let $p \in J^{+}(\Sigma)$. Since $\Sigma\subset J^+(V)$, we have that $J^-(p)\cap \Sigma\subset J^+(J^-(p)\cap V)$. In particular, $J^-(p)\cap \Sigma$ is a closed subset of the compact set $J^+(J^-(p)\cap V)\cap J^-(p)$ (note that $J^-(p)\cap V$ is compact by global hyperbolicity), and so it is also compact, that is, $\Sigma$ is future causally complete.

%$J^{-}(p) \cap \Sigma$ is a closed subset of $J^{-}(p) \cap V$ (because $\Sigma$ is closed in $V$). Since $V$ is a Cauchy hypersurface, $J^{-}(p) \cap V$ is compact, and thus, $\Sigma$ is future causally complete.

Consider the two smooth vector fields $K_{\pm}: \Sigma \rightarrow TM$ given by
\[
K_{-} := -\frac{\partial}{\partial v} \mbox{ and } K_{+} := \frac{\partial}{\partial u} - \frac{1}{2} H(x,u_0) \frac{\partial}{\partial v}.
\]
These fields are clearly null, future-directed and $g(K_{+},K_{-}) = -1$. Moreover, they are everywhere normal to $\Sigma$. In particular, $\Sigma$ is two-sided. Since $\Sigma$ is future causally complete, $J^{+}(\Sigma)$ is closed (cf. Lemma \ref{lemmamain2}). Hence, we can define subsets ${\cal H}_{\pm}$ of $\partial I^{+}(\Sigma) \setminus \Sigma$ as follows. Given a point $p \in \partial I^{+}(\Sigma) \setminus \Sigma\subset J^+(\Sigma)$, we have a null geodesic segment $\alpha:[0,1] \rightarrow \partial I^{+}(\Sigma)$ with $\alpha(1)=p$ and $\alpha(0) \in \Sigma$. Clearly, $\alpha '(0) \perp \Sigma$, and thus $\alpha '(0)$ is parallel to either $K_{+}$, in which case we say that $p \in {\cal H}_{+}$, or $K_{-}$, in which case $p \in {\cal H}_{-}$. In particular, $\partial I^{+}(\Sigma) \setminus \Sigma = {\cal H}_{-}\cup {\cal H}_{+}$. We shall refer to normal future-directed null geodesic starting at $\Sigma$ in the direction $K_{+}$ [resp. $K_{-}$] as {\em transversal} [resp. {\em longitudinal}].

We now proceed through a series of claims.

{\em Claim 1:} $\Sigma$ is acausal.

Since the hypersurface $\S_0$ is achronal, then any future-directed causal curve segment $\gamma: t \in [0,1] \mapsto (\gamma_0(t), v(t),u(t)) \in M_0 \times \mathbb{R}^2$ with $\gamma(0), \gamma(1) \in \Sigma\subset \S_0$ must actually be (up to reparametrization) a null geodesic. However, in that case our previous discussion on geodesics show that one must have $u'(t) = const.$, so if $\gamma$ is to return to $S_0$, one must have $u(t) \equiv u_0$. However, in this case, $v(t) = at + v_0$, where $a \in \mathbb{R}$. But then $a=0$ and the corresponding geodesic is spacelike, a contradiction.

{\em Claim 2:} The sets ${\cal H}_{\pm}$ are connected.

We prove this for ${\cal H}_{+}$, the other case being analogous. Let $p,q \in {\cal H}_{+}$, and let $\alpha_P , \alpha_F:[0,1] \rightarrow \partial I^{+}(\Sigma)$ past- and future-directed null geodesic segments, respectively, with $\alpha_P(1), \alpha_F(0) \in \Sigma$, $\alpha_P(0) = p$ and $\alpha_F(1) = q$, and with $\alpha_P'(1)$ and $\alpha_F'(0)$ parallel to $K_{+}$. Now, $M_0$ is path-connected, since $M$ is, and it follows that $\Sigma$ is path-connected. Pick a continuous path $\beta:[0,1] \rightarrow \Sigma$ with $\beta(0) = \alpha_P(1)$ and $\beta(1) = \alpha_F(0)$. Let $U \subseteq M$ be a connected normal neighbourhood of $\Sigma$. Then, for some connected neighbourhood $\tilde{U} \subseteq N\Sigma$ of the zero section in $N\Sigma$, the normal exponential map $\exp ^{\perp}:\tilde{U} \rightarrow U$ is a diffeomorphism. Since $\beta[0,1] \subset \Sigma$ is compact, there exists a number $0<\epsilon <1$ for which $s  K_{+}(\beta(t)) \in \tilde{U}$, for all $s \in [0,\epsilon]$, and all $t \in [0,1]$. Since $\alpha_P$ and $\alpha_F$ are maximal near $\Sigma$, we can also assume that $\alpha_P(s) = \exp ^{\perp}((1-s)K_{+}(\beta(0)))$, for all $s \in [1- \epsilon, 1]$, and  $\alpha_F(s) = \exp^{\perp}(sK_{+}(\beta(1)))$, for all $s \in [0, \epsilon]$, that is, the endpoints of $\beta$ can move along $\alpha_P$ and $\alpha_F$. By juxtaposing $\alpha_P|_{[0,1- \epsilon]}$, $\exp ^{\perp}(\epsilon K_{+}\circ \beta)$ and $\alpha_F|_{[\epsilon, 1]}$, we end up with a continuous curve in ${\cal H}_{+}$ connecting $p$ and $q$.

{\em Claim 3:} $S_{\pm} = {\cal H}_{\pm}$ and ${\cal H}_{-} \cap {\cal H}_{+} = \emptyset$.

Clearly, $S_{-} \subseteq {\cal H}_{-}$. We know that {\em every} geodesic $\gamma(t) = (\gamma_0(t), v(t),u(t)) \in M_0 \times \mathbb{R}^2$ satisfies $u' = const.$, which applied to future-directed null longitudinal geodesics implies that ${\cal H}_{-} \subseteq S_{-}$ (because, in this case, $u'\equiv 0$). For the same reason, transversal null geodesics (i.e. $u'\neq 0$) can never return to $S_{-}$, so ${\cal H}_{-} \cap {\cal H}_{+} = \emptyset$, and $\eta(0,\infty) \cap S_{-} = \emptyset$. Finally, since ${\cal H}_{+}$ is connected and it contains $\eta(0,\infty)$, it coincides with the connected component of $\partial J^{+}(\Sigma)\setminus \Sigma$ containing $\eta(0,\infty)$.

Applying now Theorem \ref{main2} to $A = \eta(0,\infty)$ (cf. Propositions \ref{Ex1.3} and \ref{prop1.4}), we conclude that $S_{+}$ is a smooth totally geodesic null hypersurface with future-complete null generators.

{\em Claim 4:} $\S$ is a future Cauchy hypersurface.

  In order to show that $D^+(\S)=J^+(\S)$, we will focus on the inclusion to the left, as the other one is trivial. Take any $p\in J^+(\S)=J^+(\Sigma)$ and consider a nonspacelike past-directed past-inextendible curve $\alpha:[0,a)\rightarrow M$ with $\alpha(0)=p$. Assume by contradiction that $\alpha([0,a))\cap \S=\emptyset$. Then, $\alpha([0,a))\subset J^+(\Sigma)\cap J^-(p)=J^+(\Sigma\cap J^-(p))\cap J^-(p)$, which is compact because so is $\Sigma\cap J^-(p)$, in contradiction with strong causality.

 %We will focus on the inclusion to the left, as the inclusion to the right is trivial. So, let $p\in J^+(\S)=J^+(\Sigma)$ (note that $\S=\partial J^+(\Sigma)\subset J^+(\Sigma)$). Then, there exists $q\in\Sigma$ such that $q\leq p$. Let $\tilde{V}$ be a Cauchy hypersurface with $q\in\tilde{V}$. Since $\tilde{V}$ is achronal, $\tilde{V}\cap I^+(\Sigma)=\emptyset$. Let $\alpha:[0,a)\rightarrow M$ be a past-directed non-spacelike inextendible curve with $\alpha(0)=p$. Since $p\in J^+(\Sigma)$, necessarily $p\in J^+(\tilde{V})$. Hence, $\alpha\cap \tilde{V}\neq\emptyset$. %So, there exists $t_0\in [0,a)$ such that $\alpha(t_0)\in\tilde{V}$.
 %But, $J^+(\Sigma)=\overline{I^+(\Sigma)}$, $\tilde{V}\cap I^+(\Sigma)=\emptyset$ and $\alpha\not\subset I^+(\Sigma)$, so $\alpha$ must pass through $\partial J^+(\Sigma)=\S$.

%Indeed, note that the achronality of $V$ implies that $V \cap I^{+}(\Sigma) = \emptyset$. Given some $p \in J^{+}(\S)$, any past-directed, past-inextendible causal curve must intersect $V$, and hence $\S$. So $J^{+}(\S)= D^{+}(\S)$.

Finally, for the last assertion, recall that the Weingarten map $b$ vanishes identically along each null geodesic generator $\gamma$ of $S_{+}$. Then, Eq. (\ref{ricattiequation}) gives that $\langle R(v,\gamma ')\gamma ', v \rangle = 0$ for every vector $v$ normal to $\gamma '$ and tangent to $S_+$ (because $R(v,\gamma ')\gamma '$ is parallel to $\gamma'$). A direct computation shows that for every vector $v$ normal to $\gamma'$ and tangent to $S_+$, the equality
\begin{equation}\label{nueva}
0= \langle R(v,\gamma ')\gamma ', v \rangle = - \frac{1}{2}Hess_0H(v,v).
\end{equation}
holds. Now, observe that since $\partial /\partial v$ is a Killing vector field, the assumption $(ii)$ actually implies that we a have a $\Sigma_{v,u_0}$-ray for {\em every} $v \in \mathbb{R}$. This means, in turn, that the region $u>u_0$ is foliated by totally geodesic null hypersurfaces where $Hess _0 H$ vanishes, and thus $Hess _0 H =0$ therein.
\qcd

For the following corollaries, we shall fix $\Sigma \subset M$ as in Theorem \ref{main3}.

\begin{corollary}\label{cor} Under the hypothesis of previous theorem, any analytic pp-wave (that is, with $H$ analytic) is isometric to Minkowski spacetime. \end{corollary}
{\em Proof.} From the last assertion in the previous theorem, we can assume that $Hess _0H$ vanishes on the region $u>u_0$. But since $Hess _0H$ is analytic, this means that $Hess_0 H$ vanishes everywhere. Taking into account that now $(R_{g_0})^i_{jkl} \equiv 0$, we deduce that the curvature of $(M,g)$ vanishes identically, and so, $(M,g)$ is simply Minkowski spacetime.
\qcd

\begin{corollary} Assume that $(M,g)$ is a globally hyperbolic, future null geodesically complete pp-wave obeying the null convergence condition. If there exists a future-directed null $\Sigma$-line $\eta:(-\infty,\infty) \rightarrow M$ wrt $\Sigma$\footnote{I.e., $\eta$ intersects $\Sigma$ and $\eta\mid_{[0,\infty)}$ (resp. $\eta\mid_{(-\infty,0]}$) is a future-directed (resp. past-directed) null $\Sigma$-ray wrt $\Sigma$.} and not contained in the hypersurface $u = u_0$, and $\Sigma$ is contained in the causal future of a Cauchy hypersurface $V$, then $(M,g)$ is isometric to Minkowski spacetime.
\end{corollary}
{\em Proof.} Reasoning as in the proof of previous result we deduce that the region $u >u_0$ is flat. Reasoning time-dually we also deduce that the region $u <u_0$ is flat. Therefore, we can assume that $H=0$ in these regions, and so, it must vanish identically on the whole space. \qcd

\begin{corollary} Assume that $(M,g)$ is a globally hyperbolic, future null geodesically complete homogeneous (i.e., with $H$ independent of $u$) pp-wave obeying the null convergence condition. If there exists a future-directed null $\Sigma$-ray $\eta:[0,\infty) \rightarrow M$ starting at $\Sigma$ and not contained in the hypersurface $u = u_0$, then $(M,g)$ is isometric to Minkowski spacetime.
\end{corollary}
{\em Proof.} Since the metric coefficients are independent of $u$, we only need to prove that the region $u>u_0$ is flat. To this aim, and arguing as in the proof of Corollary \ref{cor}, it suffices to prove that $\Sigma$ is FCC (recall footnote 7). So, let $q=(x,u,v)\in J^{+}(\Sigma)$ and assume by contradiction that $J^-(q)\cap\Sigma$ is not compact. Then, there exists a family $(z_i)_{i\in I}$ of past-directed causal geodesics in $M$ with $z_i(0)=q$, $z_i(1)\in \Sigma$ for all $i\in I$, such that $(z_i(1))_{i\in I}$ is not contained in any compact subset of $\Sigma$. In particular, each $z_i(s)=(x_i(s),u_i(s),v_i(s))$ satisfies
\[
\begin{array}{l}
u_i(s)=\Delta u \cdot s + u_0,\qquad \Delta u=u-u_0,
\\
v_i(s) = v + \frac{1}{2 \Delta u} \int_{0}^s \left( E_i -
g_0(\dot x_i(\sigma), \dot x_i(\sigma)) + 2
V_{\Delta}(x_i(\sigma))\right) d\sigma,\qquad g(\dot{z}_i(s),\dot{z}_i(s))\equiv E_i.
\end{array}
\]
Consider the family of past-directed causal curves $\overline{z}_i(s)=(\overline{x}_i(s),\overline{u}_i(s),\overline{v}_i(s))$, $i\in I$, given by:
\[
\begin{array}{l}
\overline{x}_i(s)=\left\{\begin{array}{ll} x_i(2s) & s\in [0,1/2] \\ x_i(2-2s) & s\in [1/2,1]
\end{array}\right.
\\
\overline{u}_i(s)=\overline{\Delta u}\cdot s + u_0,\qquad \overline{\Delta u}=2\Delta u
\\
\overline{v}_i(s) = v + \frac{1}{2 \overline{\Delta u}} \int_{0}^s \left( \overline{E}_i -
g_0(\dot{\overline{x}}_i(\sigma), \dot{\overline{x}}_i(\sigma)) + 2
V_{\overline{\Delta}}(\overline{x}_i(\sigma))\right) d\sigma, \qquad \overline{E}_i=4 E_i.
\end{array}
\]
With this choice, note that $\overline{\gamma}_i(1)=(x,u_0+2\Delta u,2v_0-v)$ for all $i\in I$. Therefore, the causal diamond $J(p,q)=J^+(p)\cap J^-(q)$, with $p=(x,u_0+2\Delta u,2v_0-v)$, satisfies that $J(p,q)\cap \Sigma\supset (z_i(1))_{i\in I}$ is not contained in any compact set of $\Sigma$, and so, $J(p,q)$ cannot be compact, in contradiction with global hyperbolicity. \qcd

\begin{remark}
\label{ray}
{\em In all these rigidity results we have adopted the existence of a null $\Sigma$-ray as one of our hypotheses. This assumption is natural, for example, in the following setting. Suppose that $(M,g)$ is a generalized plane wave spacetime admitting a suitably regular, conformal future infinity ${\cal J}^{+}$, and that our surface $\Sigma$ (defined again as in Theorem \ref{main3}) is {\em visible} from ${\cal J}^{+}$ in the sense that ${\cal J}^{+} \cap J^{+}(\Sigma, M \cup {\cal J}^{+}) \neq \emptyset$. Assume, furthermore, that at least one connected component of ${\cal J}^{+}$ that intersects $J^{+}(\Sigma, M \cup {\cal J}^{+})$ is not entirely contained in the latter set. (This will be the case, e.g., if ${\cal J}^{+}$ is a null hypersurface with null generators having a past endpoint in a suitably defined ``spatial infinity'' $i^0$; cf. the proof of Proposition 9.2.1 in \cite{HE}.) Then that connected component will intersect $\partial I^{+}(\Sigma,M \cup {\cal J}^{+})$, and hence there would exist a future-complete, future-inextendible null generator of $\partial I^{+}(\Sigma)$ which would have a past endpoint on $\Sigma$ (since $J^{+}(\Sigma)$ is closed). This would be our $\Sigma$-ray. This reasoning shows that the existence of such a ray will fail to hold only if the structure of $(M,g)$   is bad enough, or if $\Sigma$ is not visible from infinity, and hence within a black hole region.}
\end{remark}

\section{Bifurcate horizons around maximal surfaces}\label{5}

Suppose that $(M,g)$ admits a Killing vector field $X$ and let $f:= \langle X,X \rangle$. Consider the open set
\[
{\cal N} := \{ p \in M \, | \, X(p), \nabla f(p) \neq 0 \}\subset M.
\]
Assume that $\tilde{{\cal H}} = {\cal N}\cap f^{-1}(0)$ is non-empty. Then it is a smooth hypersurface in $(M,g)$ restricted to which $X$ is null. Using the fact that $X$ is Killing, it is easy to check that
\[
\nabla f = - 2 \nabla _X X \mbox{ and } X \langle X, X \rangle = 2 \langle X, \nabla _X X \rangle = 0,
\]
and therefore $\langle \nabla f , X \rangle \equiv 0$. Hence, $X$ is everywhere tangent to $\tilde {\cal H}$ and at any $p \in \tilde{{\cal H}}$, $\nabla f(p)$ is either null or spacelike. Any connected component of $\tilde {\cal H}$ restricted to which $\nabla f$ is everywhere null is therefore a null hypersurface, called a {\em (non-degenerate) Killing horizon} in $(M,g)$\footnote{This definition is slightly less general than some others appearing in the literature (cf. e.g., \cite{boyer,wald,chrusciel}), but since we use it only to motivate the main theorem in this section, it is enough for our purposes.}.

Let ${\cal H} \subset M$ be a Killing horizon. Since ${\cal H}$ is a null hypersurface, given $p \in {\cal H}$ and $v,w \in T_p{\cal H}$, the symmetry of the null second fundamental form (cf. Prop. \ref{prop2.2}) gives $B_p(\overline{v}, \overline{w}) = \langle v, \nabla _w X (p)\rangle = \langle w, \nabla _v X (p)\rangle$, while the fact that $X$ is Killing gives $\langle v, \nabla _w X (p)\rangle = - \langle w, \nabla _v X (p)\rangle$, and hence $B\equiv 0$ at $p$. Therefore, ${\cal H}$ is totally geodesic.

Now, let ${\cal Z}$ be the set of zeros of a Killing vector field $X$. A non-empty connected component $\Sigma$ of ${\cal Z}$ is a {\em bifurcation surface} for $X$ in $(M,g)$ if it is a compact, codimension 2, two-sided acausal (hence spacelike) submanifold. Note that this implies that $\Sigma$ is a totally geodesic submanifold of $(M,g)$ (see, e.g., Theorem 1.7.12, p. 48 of Ref. \cite{oneill2}).  It is well known (see, e.g, \cite{boyer, wald}) that if $\Sigma$ is one such bifurcation surface, there exist exactly two smooth connected null hypersurfaces ${\cal H}_1$ and ${\cal H}_2$ such that (i) $\Sigma = {\cal H}_1 \cap {\cal H}_2$, (ii) the null generators of ${\cal H}_1$, ${\cal H}_2$  at $\Sigma$ are normal to $\Sigma$ and (iii) ${\cal H}_1 \cup {\cal H}_2 \setminus \Sigma$ is the disjoint union of four Killing horizons. The union ${\cal H}_1 \cup {\cal H}_2$ is then called a {\em bifurcate Killing horizon}.

The importance of bifurcate Killing horizons is that they appear in many exact solutions of the Einstein field equation in General Relativity, for instance as event horizons in the black hole spacetimes of the Kerr-Newman family or as the cosmological horizon in de Sitter spacetime. Indeed, it has been shown that stationary black hole event horizons are portions of bifurcate Killing horizons under much more general circumstances (see. e.g., \cite{racz}). They also are the natural arena for the the study of thermal properties of quantum states in semiclassical quantum gravity, such as the Hawking radiation of black holes and the Unruh effect \cite{wald}.

We now proceed to show, as a second application of our previous results, that a similar bifurcate structure appears around maximal codimension 2 compact acausal submanifolds in certain spacetimes {\em not necessarily} possessing Killing vector fields. Specifically, we have the following

\begin{theorem}
\label{bifurcate}
Suppose that $(M,g)$ is a globally hyperbolic, null geodesically complete spacetime satisfying the null convergence condition. Suppose that there exists a compact surface $\Sigma$ such that
\begin{itemize}
\item[i)] For some Cauchy hypersurface $V$ of $(M,g)$, $\Sigma \subset V$, and $V \setminus \Sigma$ is the disjoint union of two connected $C^0$ hypersurfaces $V_{i}$, $i=1,2$, with non-compact closure,
\item[ii)] $\Sigma$ is maximal in $(M,g)$.
\end{itemize}
Then $\partial I^{+}(\Sigma)\setminus \Sigma$ [resp. $\partial I^{-}(\Sigma)\setminus \Sigma$] has exactly two connected components ${\cal H}^{+}_i$ ($i=1,2$) [resp. ${\cal H}^{-}_i$ ($i=1,2$)] homeomorphic respectively to $V_{i}$ ($i=1,2$), so that $\Sigma = \cap _{i=1}^2 \overline{{\cal H}^{+}_i} \cap \overline{{\cal H}^{-}_i}$. Moreover, these connected components are smooth and totally geodesic null hypersurfaces in $(M,g)$ with future-complete [resp. past-complete] null generators.
\end{theorem}
{\em Proof.} One can focus on the analysis of $\partial I^{+}(\Sigma)\setminus \Sigma$, since the arguments for $\partial I^{-}(\Sigma)\setminus \Sigma$ are time-dual. Now, recall that, fixing a timelike vector field $X: M \rightarrow TM$, its maximally extended integral curves define a continuous open onto mapping $\rho _X: M \rightarrow V$ leaving $V$ pointwise fixed (cf. Proposition 14.31 of \cite{oneill}). By arguing just as in the proof of Proposition 2.4 of \cite{me2} and using (i), we then show that $\partial I^{+}(\Sigma)\setminus \Sigma$ has exactly two connected components ${\cal H}^{+}_i$ ($i=1,2$) and that the restrictions of $\rho _X$ to ${\cal H}^{+}_i$ induce a homeomorphism from each of these null hypersurfaces onto one of the hypersurfaces $V_{i}$, $i=1,2$. This establishes the first part. To prove the second part, we only need to deal with one of the components, say ${\cal H}^{+}_1$, since the arguments for the other are the same. Now, fix a complete Riemannian metric $h$ on $M$. Since the closure of ${\cal H}^{+}_1$ is non-compact, there exists a sequence of points $(p_n) \subseteq {\cal H}^{+}_1$ such that for each compact $K \subseteq M$, $p_n \not\in K$ for all large enough $n$. Therefore, for each $n \in \mathbb{N}$, we can pick a future-directed nonspacelike curve $\gamma_n:[0, +\infty) \rightarrow M$ parametrized by h-arc length such that for some $t_n \in (0, +\infty)$, $\gamma_n(t_n) = p_n$ and $\gamma_n|_{[0,t_n]}$ is a null pregeodesic normal to $\Sigma$ and without focal points before $p_n$. Using the compactness of $\Sigma$, we can use a standard limit curve argument (cf., e.g., the proof of Proposition 2.1 of \cite{me2}) to establish that there exists a future-directed, affinely parametrized null $\Sigma$-ray $\gamma:[0, +\infty) \rightarrow M$ entirely contained in ${\cal H}^{+}_1$. The fact that ${\cal H}^{+}_1$ is smooth and totally geodesic now follows from $(ii)$ by an application of Theorem \ref{main2}.
\qcd

Generalizing the above definitions, suppose that there exist two connected null hypersurfaces ${\cal H}_1$ and ${\cal H}_2$ and a compact surface $\Sigma \subseteq M$ such that (i) $\Sigma = {\cal H}_1 \cap {\cal H}_2$, (ii) the null generators of ${\cal H}_1$, ${\cal H}_2$  at $\Sigma$ are normal to $\Sigma$ and (iii) ${\cal H}_1 \cup {\cal H}_2 \setminus \Sigma$ is the disjoint union of four connected totally geodesic null hypersurfaces. Then, we call the union ${\cal H}_1 \cup {\cal H}_2$ a {\em bifurcate horizon} in $(M,g)$.

We now proceed to consider a natural setting in which the assumptions in Theorem \ref{bifurcate} are expected to hold. We shall not strive here for maximum generality, but will be contended with the simplest situation in which this occurs. Let us first recall some fairly standard definitions .

By an {\em ($n$-dimensional) initial data set} we mean a triple $(N,h,K)$, where $(N,h)$ is a smooth $n$-dimensional Riemannian manifold and $K$ is a smooth, symmetric $(0,2)$ tensor field over $N$. We can then define a real-valued function $\rho$ and a 1-form $J$ on $N$ by
\begin{eqnarray}
\label{constraint1}
\rho &:=& \frac{1}{2} \left( R_N - |K|^2 + \left( \tr _N K \right) ^2 \right); \\
\label{constraint2} J &:=& {\mbox div} _N \left( K - \left( \tr _N K \right) h \right),
\end{eqnarray}
where $R_N$ denotes the scalar curvature in $N$. The initial data set $(N,h,K)$  is {\em vacuum} if $\rho$ and $J$ for this set vanish identically, and {\em time-symmetric} if $K=0$.

Of course, the definitions above are purely geometric, but they acquire physical importance when applied in the initial-value formulation of General Relativity. In this setting, $N$ is to be thought of as an embedded spacelike hypersurface in an $(n+1)$-dimensional spacetime $(M,g)$, with $h$ being the induced metric and $K$ being the second fundamental form, taken with respect to the unique future-directed, unit timelike normal vector field $U$ over $N \hookrightarrow M$. Moreover, $(M,g)$ is assumed to satisfy the Einstein field equation
\begin{equation}
\label{EFE}
Ric_M - \frac{1}{2} R_M g + \Lambda g = T,
\end{equation}
for a suitable $(0,2)$ symmetric energy-momentum tensor $T$, and a (possibly vanishing) cosmological constant $\Lambda$. One can check that the Gauss-Codazzi equations for the embedding $N \hookrightarrow M$ imply that the initial data $(N,h,K)$ automatically satisfies Eqs. (\ref{constraint1}) and (\ref{constraint2}) with the identifications
\begin{eqnarray}
\rho &\equiv & T(U,U) + \Lambda, \\
J &\equiv & -T(U, \cdot ).
\end{eqnarray}
If in addition $(M,g)$ is globally hyperbolic and $N$ is a Cauchy hypersurface therein, then we say that $(M,g)$ is a {\em Cauchy development} of the initial data set $(N,h,K)$. It is well-known \cite{choquetbruhat} that for any vacuum initial data, there exists a unique Cauchy development (with $T=0$ and $\Lambda =0$) which is inextendible in the class of globally hyperbolic spacetimes. We shall refer to this as {\em the} Cauchy development of the corresponding data.

For simplicity, in what follows we shall only be interested in {\em time-symmetric and vacuum} initial data. This reduces (by Eqs. \ref{constraint1} and \ref{constraint2}) simply to a Riemannian manifold $(N,h)$ with (identically) zero scalar curvature. Our motivation now, as mentioned, is just to see that the assumptions of Theorem \ref{bifurcate} are naturally met in some concrete settings. One especially important such setting is when $(N,h)$ is asymptotically flat. This concept of course arises quite naturally in General Relativity, to describe gravitationally isolated systems. Let us first define more precisely what we mean here by `asymptotically flat'.

\begin{definition}
\label{AFdef}
$(N,h)$ is {\em asymptotically flat} if it is complete and there exists a compact subset $\Omega \subset N$ such that $N\setminus \Omega$ has a finite number of components $E_1, \ldots, E_k$, called {\em (asymptotically flat) ends}, and each such end $E_i$ is diffeomorphic to the region $\{ x \in \mathbb{R}^n \; : \; |x| >1 \}$. In addition, the metric components in the coordinate system induced by this diffeomorphism satisfy the estimates
\[
|h_{ij}(x) - \delta_{ij}| \leq \frac{C}{|x|^{\alpha}}\; \mbox{     and     }
\; |\partial _k h_{ij}(x)| \leq \frac{C}{|x|^{\alpha +1}},
\]
for some positive constants $C,\alpha$. \footnote{The definition of asymptotic flatness varies slightly throughout the literature. The one presented here is enough for our purposes.}
\end{definition}

We are now ready to state the following result.
\begin{theorem}
\label{reasonable}
Let $(N,h)$ be an asymptotically flat Riemannian manifold of dimension $3 \leq n \leq 7$ with two asymptotically flat ends and zero scalar curvature. Suppose that $(M,g)$ is the vacuum Cauchy development of $(N,h)$ (viewed as a vacuum initial data set). Then, either $(M,g)$ is null geodesically incomplete or it admits a bifurcate horizon.
\end{theorem}
{\em Proof.} Let $E_{+} \subset N$ be either one of the asymptotically flat ends, and let
\[
\psi : R=\{ x \in \mathbb{R}^n \; : \; |x| >1 \} \rightarrow E_+
\]
be a diffeomorphism. Let $r>1$ and $S_r = \{ x \in \mathbb{R}^n \; : \; |x| = r \} \subset R$. The image of $S_r$ by $\psi$ (which by a slight abuse of notation we also denote by $S_r$ in what follows) is an embedded hypersurface contained in $E_+$. If we write
\[
N_{+} = \psi\left(\{ x \in \mathbb{R}^n \; : \; |x| >r \}\right), \\
N_{-} = N \setminus \psi\left(\{ x \in \mathbb{R}^n \; : \; |x| \geq r \}\right),
\]
then, clearly, $S_r \cup N_{+}$ is noncompact and $i_{\#}:\pi_1(S_r) \rightarrow \pi_1(S_r \cup N_{+})$ induced by the inclusion is surjective. It is straightforward to check that for some choice of $r>0$, $S_r$ has positive mean curvature with respect to the unit normal pointing into $N_{+}$, independently of the specific decay constants $C$ and $\alpha$.

The same construction of course applies to the other end $E_{-}$ as well, and we end up with two mean convex spheres $S_{\pm}$ which together form the boundary of a compact $n$-dimensional submanifold $\Omega$ such that $N = N_{+} \cup N_{-} \cup \Omega$. The conditions of Theorem 1.1 in \cite{eichmair2} on the existence of MOTS are now satisfied  (cf. also Theorem 3.3 of Ref. \cite{AEM}), hence there exists a minimal (i.e., having zero mean curvature \footnote{Recall that a MOTS in a time-symmetric initial data is just a minimal surface therein.}) compact connected hypersurface $\Sigma \subset N$ in the interior of $\Omega$ which separates $\Omega$.

Now, inside the vacuum Cauchy development $(M,g)$ of $(N,h)$, $N$ is an achronal spacelike hypersurface, and hence acausal (cf. Theorem 42, p. 425 of \cite{oneill}). Thus, $\Sigma$ clearly satisfies all the conditions of Theorem \ref{bifurcate} if $(M,g)$ is geodesically complete, and the result now follows.

\qcd

\section{Acknowledgements}
IPCS is partially supported by FAPESC Grant 2014TR2954.
JLF is partially supported by the Spanish MICINN-FEDER Grant MTM2010-18099 and the Regional J. Andaluc\'{i}a Grant PP09-FQM-4496, with FEDER funds. He gratefully acknowledges the hospitality of the Department of Mathematics of Universidade Federal de Santa Catarina.

%%%%%%%%%%%%%%%%%%%%%%%%%%%%%%%%%%%%%%%%%%%%%%%%%%%%%%%%%%%%%%%%%%%%%%%%%%%%%%%%%%%%%
%References
%%%%%%%%%%%%%%%%%%%%%%%%%%%%%%%%%%%%%%%%%%%%%%%%%%%%%%%%%%%%%%%%%%%%%%%%%%%%%%%%%%%%%

\vspace*{\fill}

%-------------------------------------------------------------------------------------------------------------

\vspace{.5cm}


\begin{thebibliography}{1}
%
\bibitem{AEM}
L. Andersson, M. Eichmair and J. Metzger, {\em Jang's equation and its applications to marginally trapped surfaces}, in Complex Analysis and Dynamical Systems IV: Part 2. General Relativity, Geometry and PDE, Contemporary Mathematics, vol. 554, AMS and Bar-Ilan (2011).
%
\bibitem{AMMS}
L. Andersson, M. Mars, J. Metzger and W. Simon, {\em The time evolution of marginally trapped surfaces}, Class. Quantum Grav. {\bf 26} (2009), 085018.
%
\bibitem{AM}
L. Andersson and J. Metzger, {\em The area of horizons and the trapped region}, Comm. Math. Phys. {\bf 290} (2009) 941-972.
%
\bibitem{BE}
J.K. Beem, P.E. Ehrlich and K.L. Easley
  {\it Global Lorentzian Geometry}, $2^{\rm{nd}}$ ed.,
  Marcel Dekker, New York (1996).
%
\bibitem{boyer}
R.H. Boyer, {\em Geodesic Killing orbits and bifurcate Killing horizons}, Proc. Roy. Soc. London {\bf A311} (1969) 245-252.
%

\bibitem{Bi} J. Bicak: {\it Selected solutions of Einstein's field equations: Their role in general relativity and astrophysics.} In: {\it Einstein's Field Equations and Their Physical Interpretations,} Lect. Notes Phys. \textbf{540}, ed by B. Schmidt (Springer, Heidelberg 2000) pp 1-126. Available at gr-qc/0004016

\bibitem{Brinkmann} H. Brinkmann, {\em Einstein spaces which are conformally mapped on each other}, Math. Ann. {\bf 94} (1925) 119-145.

\bibitem{CFSgrg} A.M. Candela, J.L. Flores, M. S\'anchez: Gen. Rel. Grav. \textbf{35}, 631 (2003). Available at gr-qc/0211017.

%
\bibitem{choquetbruhat}
Y. Choquet-Bruhat and R. Geroch, {\em Global aspects of the Cauchy problem in general relativity}, Commun. Math. Phys. {\bf 14} (1969) 329-335.
%
\bibitem{chrusciel}
P.T. Chrusciel, {\em ``No Hair'' Theorems - folklore, conjectures, results}, Differential Geometry and Mathematical Physics (J. Beem and K.L. Duggal, eds.), Cont. Math. vol. 170, AMS, Providence (1994) 23-49.
%
\bibitem{CG}
P.T. Chrusciel and G.J. Galloway, {\em Outrer trapped surfaces are dense near MOTSs},  arXiv:1308.6057v2.
%
\bibitem{me2}
I.P. Costa e Silva, {\em On the geodesic incompleteness of spacetimes containing marginally (outer) trapped surfaces}, Class. Quantum Grav. {\bf 29} (2012) 235008.
%
\bibitem{EK}
J. Ehlers and K. Kundt, {\em Exact solutions of the gravitational field equations}, in {\em Gravitation: a introduction to current research}, L. Witten (ed.), J. Wiley \& Sons, New York (1962) pp. 49-101.
%
\bibitem{eichmair1}
M. Eichmair, {\em The Plateau problem for marginally outer trapped surfaces}, J. Diff. Geom. {\bf 83} (2009) 551-583.
%
\bibitem{eichmair2}
M. Eichmair, {\em Existence, regularity, and properties of generalized apparent horizons}, Comm. Math. Phys. {\bf 294} (2010) 745-760.
%
\bibitem{EGP}
M. Eichmair, G.J. Galloway and D. Pollack, {\em Topological censorship from the initial data point of view}, J. Differential Geom. {\bf 95} (2013), no. 3, 389-405.
%
\bibitem{ER} A. Einstein and N. Rosen, {\em On gravitational waves}, J. Franklin Inst. {\bf 223} (1937) 43-54.
%
\bibitem{beijing}
G.J. Galloway, {\em Spacetime geometry}, Beijing lecture notes, available at  http://www.math.miami.edu/~galloway/
%
\bibitem{gallowaymaximum}
G.J. Galloway, {\em Maximum principles for null hypersurfaces and null splitting theorems}, Ann. Henri Poincaré, {\bf 1} (2000) 543-567.
%
\bibitem{harris}
S.G. Harris, {\em A triangle comparison theorem for Lorentz manifolds}, Indiana Univ. Math. J. {\bf 31} (1982) 289-308.
%
\bibitem{HE}
S.W. Hawking and G.F.R. Ellis,
  {\it The Large Scale Structure of Space-time},
  Cambridge University Press, Cambridge (1973).
%
\bibitem{wald}
B. Kay and R.M. Wald, {\em Theorems on the uniqueness and thermal properties of stationary nonsingular, quasifree states on spacetimes with a bifurcate Killing horizon}, Phys. Rep. {\bf 207} (1991), 49-136.
%
\bibitem{kupeli}
D.N. Kupeli, {\em On null submanifolds in spacetime}, Geom. Dedicata {\bf 23} (1987) 33-51.
%
\bibitem{oneill}
B. O'Neill,
  {\it Semi-Riemannian Geometry with Applications to Relativity},
  A K Peters, Wellesley, Ma. (1983).
%
\bibitem{oneill2}
B. O'Neill,
  {\it The Geometry of Kerr Black Holes},
  Academic Press, New York (1995).
%
\bibitem{P}
R. Penrose, {\em Techniques of Differential Topology }, Society for Industrial and Applied Mathematics, Philadelphia, Pa. (1972).
%
\bibitem{Penrose}
R. Penrose, {\em Gravitational collapse and space-time singularities}, Phys. Rev. Lett. {\bf 14} (1965) 57-59.
%
\bibitem{racz}
I. Racz and R.M. Wald, {\em Extensions of spacetimes with Killing horizons}, Class. Quantum Grav. {\bf 9} (1992), 2643-2656.
%
\bibitem{SY3}
R. Schoen and S.-T. Yau, {\em The existence of a black hole due to condensation of matter}, Commun. Math. Phys. {\bf 90} (1983) 575-579.
%

\bibitem{SKMHH} H. Stephani, D. Kramer, M. MacCallum, C. Hoenselaers, E. Herlt: {\it Exact Solutions of Einstein's Field Equations} (Cambridge University Press, Cambridge 2003)

\bibitem{treude}
J.H. Treude, {\em Ricci curvature comparison in Riemannian and Lorentzian geometry}, diplomarbeit, Albert-Ludwigs-Universität, Freiburg (2011).
(Available at $http://www.freidok.uni-freiburg.de/volltexte/8405/pdf/diplomarbeit_{-}treude.pdf$)
%

\bibitem{Yurtsever} U. Yurtsever, {\em Colliding almost-plane gravitational waves: colliding plane waves and general properties of almost-plane-wave spacetimes}, Phys. Rev. D {\bf 37} (1988) 2803-2817.





\end{thebibliography}
\end{document}